\documentclass[review]{elsarticle}

\usepackage{lineno}
\usepackage[colorlinks=true,linkcolor=black,citecolor=blue,urlcolor=blue,pdfauthor=author]{hyperref}
\modulolinenumbers[5]
\usepackage{amsmath,amsfonts,amssymb,amsthm,bm}
\usepackage{float}
\usepackage{graphics}
\usepackage{graphicx}
\usepackage{caption}
\graphicspath{{figures/}}
\usepackage{subfigure}
\usepackage{fullpage}
\usepackage{multirow}
\usepackage{booktabs}
\usepackage{textcomp}
\usepackage{gensymb}
\usepackage{nomencl}
\usepackage{comment}
\usepackage{setspace}
\usepackage{geometry}
\usepackage[section]{placeins}
\usepackage{soul}
\journal{Elsevier}

\makeatletter
\def\ps@pprintTitle{%
	\let\@oddhead\@empty
	\let\@evenhead\@empty
	\def\@oddfoot{}%
	\let\@evenfoot\@oddfoot}
\makeatother

\begin{document}
\captionsetup[figure]{labelfont={bf},name={Fig.},labelsep=period}
\begin{frontmatter}

\title{A multiple scattering formulation for elastic wave propagation in space-time modulated metamaterials}

\author[add1]{Xingbo Pu}
\author[add1]{Alessandro Marzani\corref{corr1}}
\ead{alessandro.marzani@unibo.it}
\cortext[corr1]{Corresponding authors}
\author[add1]{Antonio Palermo\corref{corr1}}
\ead{antonio.palermo6@unibo.it}

\address[add1]{Department of Civil, Chemical, Environmental and Materials Engineering, University of Bologna, 40136 Bologna, Italy}

\begin{abstract}

Space-time modulation of material parameters offers new possibilities for manipulating elastic wave propagation by exploiting time-reversal symmetry breaking. Here we propose and validate a general framework based on the multiple scattering theory to model space-time modulated elastic metamaterials, namely elastic waveguides equipped with modulated resonators. The formulation allows to consider an arbitrary distribution of resonators with a generic space-time modulation profile and compute the wavefield within and outside the resonators' region. Additionally, under appropriate assumptions, the same framework can be exploited to predict the waveguide dispersion relation. We demonstrate the capabilities of our formulation by revisiting the dynamics of two representative  space-time modulated systems, e.g. the non-reciprocal propagation of (i) flexural waves along a metabeam and (ii) surface acoustic waves along a metasurface. Given its flexibility, the proposed method can pave the way towards the design of novel devices able to realize unidirectional transport of elastic energy for vibration isolation, signal processing and energy harvesting purposes.

\end{abstract}


\end{frontmatter}


\section{Introduction}

In the last decade, the research on active (or activated) materials has fueled the discovery of novel dynamic functionalities to design devices for vibrations and waves control \cite{nassar2020nonreciprocity,chen2021realization}. Activated materials are often characterized by constitutive properties that are modulated in space and time according to an external energy source. The study of such space-time modulated materials was originally pioneered in optics \cite{sounas2017non} and, shortly afterward, extended to acoustics \cite{rasmussen2021acoustic} and elasticity \cite{nassar2020nonreciprocity}. Elastic waves propagating in these space-time varying media are of particular interest since the modulation can create a directional bias that breaks the time-reversal symmetry. Breaking reciprocity allows to realize rich and unconventional phenomena, including, but not limited to, unidirectional wave propagation, adiabatic energy pumping \cite{xu2020physical,riva2021adiabatic}, frequency conversion \cite{wu2022independent}. These effects can be leveraged to design novel devices such as acoustic rectifiers \cite{liang2010acoustic}, circulators \cite{fleury2014sound}, and topological insulators \cite{fleury2016floquet}, which can find  applications in acoustic communication, signal processing, energy harvesting and vibration isolation \cite{reiskarimian2016magnetic,jalvsic2023active,huang2023towards}.

In the context of elastodynamics, space-time modulation can be achieved following two strategies. The first one relies on a bias directly introduced in the waveguide, as a   modulation of the elastic and/or mass properties, so to obtain a modulated phononic crystal  \cite{trainiti2016non,nassar2017modulated,goldsberry2020nonreciprocal}. The second option utilizes space-time modulated mechanical oscillators attached to a non-modulated waveguide \cite{nassar2017EML,wu2021non,palermo2020surface} to obtain a modulated elastic metamaterial. Both approaches proved to be technically feasible by a series of experimental works where programmable electric components were used to modulate the media/oscillators \cite{chen2019nonreciprocal,trainiti2019time,marconi2020experimental,wu2022independent,wan2022low}. Nonetheless, modulated metamaterials, compared to their phononic counterpart, are  easier to realize, since only the resonant elements need to be modulated, and support non-reciprocal effects at sub-wavelength scales.

Besides the numerous examples of modulated waveguides \cite{attarzadeh2018non,chen2021efficient}, most of the conducted studies rely on the use of numerical simulations, typically developed via finite element (FE) or finite difference (FD) algorithms, to describe the expected non-reciprocal effects. Nonetheless, numerical simulations are always bounded by their computational cost which inherently limits the development of design and optimization studies. Computationally inexpensive analytical tools for modulated media are thus desirable, not only to reduce the computational burden but also to gain a deeper understanding of non-reciprocal effects. Currently, analytical formulations for time-modulated systems are mainly used to predict the dispersion relations of both  discrete \cite{vila2017bloch,nassar2017PRSA,wang2018observation} and continuous media \cite{nassar2017modulated,nassar2017EML,chen2019nonreciprocal,trainiti2019time,marconi2020experimental,wu2021non,palermo2020surface}. 

Although knowledge of the dispersion relations provides physical insights into the existence of directional band gaps, evidence of non-reciprocal phenomena can be found only by computing transient or steady-state responses across finite modulated systems. To the best of our knowledge, analytical methods for the computation of wavefields and transmission/reflection coefficients are currently limited to one-dimensional (1D) problems \cite{li2019transfer,li2019nonreciprocal,chen2019nonreciprocal,goldsberry2020nonreciprocal}. Additionally, there is no unified framework that enables the computation of both the dispersion relation and the wavefield of a generic modulated system.

To fill this gap, we here propose a generalized multiple scattering formulation able to model the dynamic response of space-time modulated resonators coupled to a generic elastic waveguide. As observed in experiments, space-time modulated resonators can generate scattered fields at lower and higher harmonics with respect to the excitation frequency \cite{chen2019nonreciprocal}. 
To capture this response, we first describe the coupling between the vibrating resonators and the waveguide motion  with an ad-hoc impedance operator able to account for the expected additional harmonics. Then, we compute the scattered fields in the waveguide  by means of Green's functions. Finally, we set our multiple scattering scheme to couple the incident and scattered fields and compute the related unknown amplitudes  by imposing proper boundary conditions at each resonator base. The proposed formulation allows us to investigate the dynamic of an arbitrary number $N$ of resonators with an arbitrary spatial-temporal modulation profile, since all the space-time varying oscillators can be described individually. Additionally, by introducing appropriate assumptions, the same formulation can be used to derive the related dispersion equations. 

The details of the methodology and its modeling capabilities are discussed in what follows. In particular, in Section \ref{Theoretical formulation} we describe the proposed general multiple scattering formulation for the computation of the wavefield and the dispersion relation of waveguides coupled with space-time-modulated resonators. In Section \ref{Examples and applications}, we apply the formulation to model flexural waves in a beam and Rayleigh waves on a substrate, both coupled with an array of modulated surface resonators. For both scenarios we show the capability of the formulation to predict non-reciprocal guided waves. Finally, we derive conclusions and outlook of the work in Section \ref{Conclusion}.

\section{Theoretical formulation} \label{Theoretical formulation}

\subsection{Statement of the problem}

\begin{figure}[htbp]
	\centering
	\includegraphics[width=0.7\textwidth]{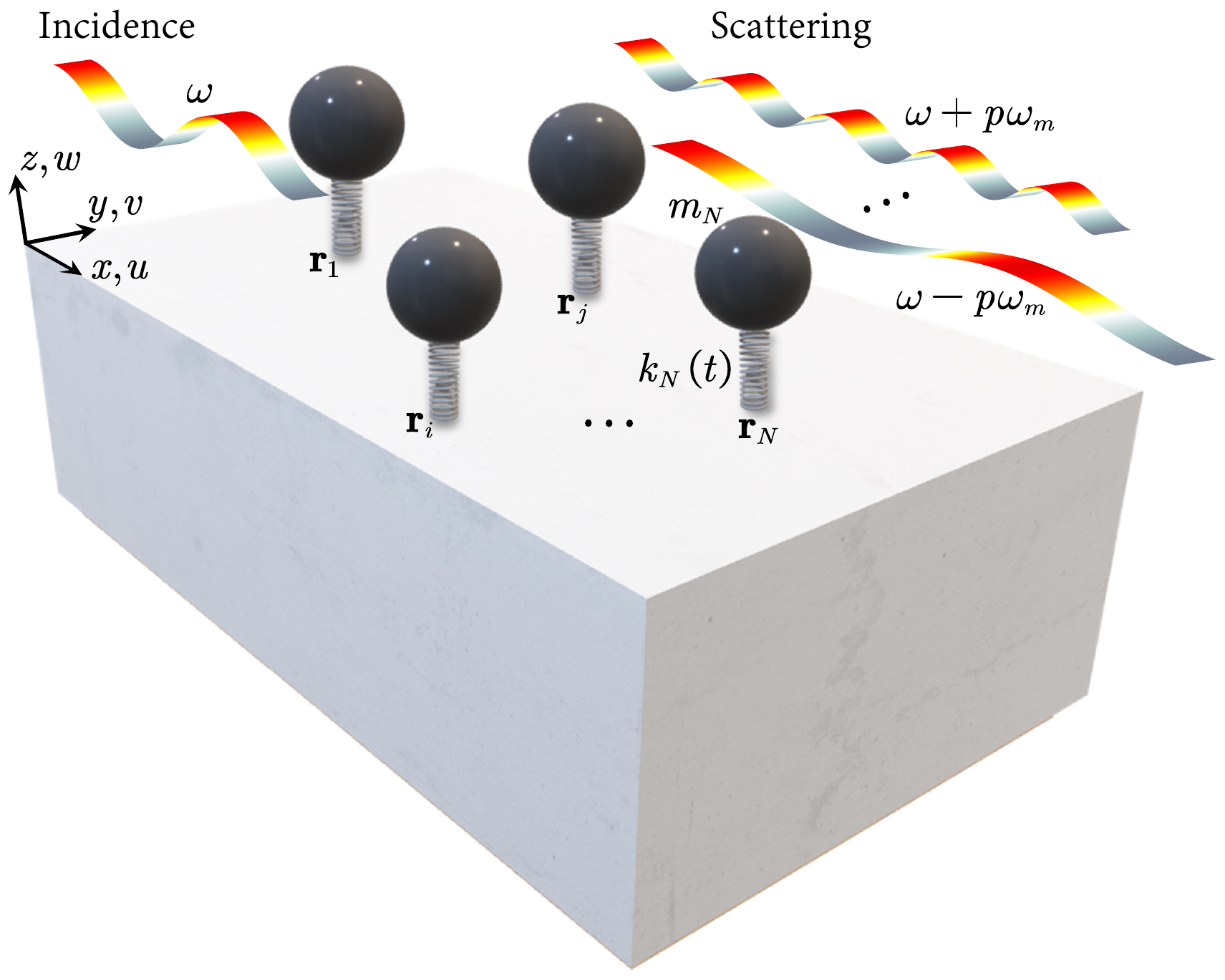}
	\caption{Schematics of space-time modulated resonators laying over an elastic waveguide.}
	\label{fig:fig1}
\end{figure}

We propose a general analytical framework to model a cluster of space-time-modulated oscillators attached to a given elastic waveguide (Fig. \ref{fig:fig1}). The formulation includes the following three steps: (i) the definition of the elastic force exerted on the waveguide by a time-modulated resonator when excited by a base motion; (ii) the use of Green's functions to describe the scattered wavefields generated by the resonators in the waveguide; (iii) the construction of a multiple scattering formulation to couple the waveguide with an arbitrary number of time-modulated resonators. The approach allows computing the lower- and higher-order scattered harmonics, generated by the collective response of the time-modulated resonators, and responsible for the non-reciprocal wave motion in the waveguide.

First, we present the framework in its most general form, i.e. considering a finite array of time-modulated oscillators with mechanical properties obeying the same modulation period $T_m$ and arbitrarily arranged over the waveguide surface. Then, we show how to derive the waveguide dispersion relation by introducing appropriate assumptions, e.g., considering an infinite array of identical resonators regularly arranged along the elastic support.

\subsection{Elastic force of a time-modulated resonator}

Let us recall the dynamics of the generic $n$th resonator attached to the waveguide surface at the location $\mathbf{r}_n$ (see Fig. \ref{fig:fig1}). The resonator has a mass $m_n$, damping coefficient $c_n$, and time-modulated spring stiffness $k_n(t)$:

\begin{equation} \label{equ:modulation function}
    k_n(t)=k_n(t+T_m),
\end{equation}
\noindent where $T_m$ is the modulation time period. 
The governing equation of the $n$th resonator motion reads:

\begin{equation} \label{equ:resonator motion}
    m_n \frac{\partial^2 W_n(t)}{\partial t^2}+c_n\left[\frac{\partial W_n(t)}{\partial t}-\frac{\partial w_n(t)}{\partial t}\right]+k_n(t)[W_n(t)-w_n(t)]=0,
\end{equation}

\noindent in which $W_n(t)=W(\mathbf{r}_n,t)$ denotes the mass vertical displacement  while $w_n(t)=w(\mathbf{r}_n,t)$ is the vertical motion at the resonator base. Accordingly, the point force $F_n(t)=F(\mathbf{r}_n,t)$ exerted by the resonator onto the waveguide surface reads:

\begin{equation} \label{equ:resonator force}
    F_n(t)=-m_n \frac{\partial^2 W_n(t)}{\partial t^2}.
\end{equation}

Since the modulated stiffness $k_n(t)$ in Eq. \eqref{equ:modulation function} is time-periodic, we express it in Fourier series form as:

\begin{equation} \label{equ:stiffness FT}
    k_n(t) = \sum_{j=-\infty}^{\infty} \hat{k}_n^{(j)}\mathrm{e}^{\mathrm{i} j\omega_m t}, \quad j\in \mathbb{Z}, 
\end{equation}
\noindent in which $\mathrm{i}=\sqrt{-1}$ is the imaginary unit, $\omega_m=2\pi/T_m$ is the modulation frequency, and where the Fourier coefficients are defined as:

\begin{equation} \label{equ:stiffness FT coeff}
    \hat{k}_n^{(j)}=\frac{\omega_m}{2\pi}\int_{\frac{-\pi}{\omega_m}}^{\frac{\pi}{\omega_m}} k_n(t) \mathrm{e}^{-\mathrm{i} j\omega_m t}\,\mathrm{d}t. 
\end{equation}
\noindent

As we will see in the next section, the motion along the waveguide excited by a harmonic ($\mathrm{e}^{\mathrm{i} \omega t}$) incident field, contains several lower- and higher-order harmonics generated by the scattering of the time-modulated mechanical resonators. As a result, the vertical motion at the resonator base, namely the motion at the waveguide surface, can be written as \cite{li2019transfer}:

\begin{equation} \label{equ:base motion FT}
    w_n(t) = \sum_{h=-\infty}^{\infty} \hat{w}_n^{(h)}\mathrm{e}^{\mathrm{i}(\omega+ h\omega_m) t}, \quad h\in \mathbb{Z}, 
\end{equation}

\noindent so that the solution of Eq. \eqref{equ:resonator motion} is sought in the form \cite{trainiti2019time,chen2019nonreciprocal,goldsberry2020nonreciprocal}:

\begin{equation} \label{equ:resonator disp FT}
    W_n(t) = \sum_{h=-\infty}^{\infty} \hat{W}_n^{(h)}\mathrm{e}^{\mathrm{i} (\omega+h\omega_m) t}, \quad h\in \mathbb{Z}.
\end{equation}

\noindent
Substituting Eqs. \eqref{equ:stiffness FT}, \eqref{equ:base motion FT} and \eqref{equ:resonator disp FT} into Eq. \eqref{equ:resonator motion}, yields:

\begin{equation} \label{equ:resonator motion FT}
\begin{split}
    & \sum_{h=-\infty}^{\infty} [-m_n(\omega+h\omega_m)^2+\mathrm{i}c_n(\omega+h\omega_m)] \hat{W}_n^{(h)}\mathrm{e}^{\mathrm{i} h\omega_m t} + \sum_{h=-\infty}^{\infty} \sum_{j=-\infty}^{\infty} \hat{k}_n^{(j)} \hat{W}_n^{(h)}\mathrm{e}^{\mathrm{i} (j+h)\omega_m t} \\
    &=\sum_{h=-\infty}^{\infty} \mathrm{i}c_n(\omega+h\omega_m) \hat{w}_n^{(h)}\mathrm{e}^{\mathrm{i} h\omega_m t} + \sum_{h=-\infty}^{\infty} \sum_{j=-\infty}^{\infty} \hat{k}_n^{(j)} \hat{w}_n^{(h)}\mathrm{e}^{\mathrm{i} (j+h)\omega_m t}.
\end{split}
\end{equation}

Exploiting the orthogonality of harmonic functions, we simplify Eq. \eqref{equ:resonator motion FT} by multiplying it for $\omega_m\mathrm{e}^{-\mathrm{i} p\omega_m t}/(2\pi)$, and integrating it from $-\pi/\omega_m$ to $\pi/\omega_m$, to obtain:

\begin{equation} \label{equ:reduced resonator motion FT}
\begin{split}
    [-m_n(\omega+p\omega_m)^2+\mathrm{i}c_n(\omega+p\omega_m)] \hat{W}_n^{(p)}+\sum_{j=-\infty}^{\infty} \hat{k}_n^{(j)} \hat{W}_n^{(p-j)}= \mathrm{i}c_n(\omega+p\omega_m)\hat{w}_n^{(p)}+\sum_{j=-\infty}^{\infty} \hat{k}_n^{(j)} \hat{w}_n^{(p-j)}, \quad p\in \mathbb{Z}.
\end{split}
\end{equation}
By truncating the orders from $-P$ to $P$, Eq. \eqref{equ:reduced resonator motion FT} can be reorganized in matrix form as:

\begin{equation} \label{equ:Mn and Qn}
    \mathbf{M}_n \mathbf{\hat{W}}_n = \mathbf{Q}_n \mathbf{\hat{w}}_n,
\end{equation}
with:

\begin{equation}
\begin{aligned}
&\mathbf{M}_n=\left[\begin{array}{rrrrr}
\hat{m}_n^{(-P)} & \hat{k}_n^{(-1)} & \hat{k}_n^{(-2)} & \cdots & \hat{k}_n^{(-2P)} \\
\hat{k}_n^{(1)} & \hat{m}_n^{(-P+1)} & \hat{k}_n^{(-1)} & \cdots & \hat{k}_n^{(-2P+1)} \\
\hat{k}_n^{(2)} & \hat{k}_n^{(1)} & \hat{m}_n^{(-P+2)} & \cdots & \hat{k}_n^{(-2P+2)} \\
\vdots & \vdots & \vdots & \ddots & \vdots \\
\hat{k}_n^{(2P)} & \hat{k}_n^{(2P-1)} & \hat{k}_n^{(2P-2)} & \cdots & \hat{m}_n^{(P)}
\end{array}\right],
\mathbf{Q}_n=\left[\begin{array}{rrrrr}
\hat{q}_n^{(-P)} & \hat{k}_n^{(-1)} & \hat{k}_n^{(-2)} & \cdots & \hat{k}_n^{(-2P)} \\
\hat{k}_n^{(1)} & \hat{q}_n^{(-P+1)} & \hat{k}_n^{(-1)} & \cdots & \hat{k}_n^{(-2P+1)} \\
\hat{k}_n^{(2)} & \hat{k}_n^{(1)} & \hat{q}_n^{(-P+2)} & \cdots & \hat{k}_n^{(-2P+2)} \\
\vdots & \vdots & \vdots & \ddots & \vdots \\
\hat{k}_n^{(2P)} & \hat{k}_n^{(2P-1)} & \hat{k}_n^{(2P-2)} & \cdots & \hat{q}_n^{(P)}
\end{array}\right], \\[2ex]
&\mathbf{\hat{W}}_n=[\hat{W}_n^{(-P)}, \hat{W}_n^{(-P+1)},\cdots, \hat{W}_n^{(P-1)},\hat{W}_n^{(P)}]^T,\quad \mathbf{\hat{w}}_n=[\hat{w}_n^{(-P)}, \hat{w}_n^{(-P+1)},\cdots, \hat{w}_n^{(P-1)},\hat{w}_n^{(P)}]^T,
\end{aligned}
\end{equation}
\noindent in which $\hat{m}_n^{(j)}=\hat{k}_n^{(0)}-m_n(\omega+j\omega_m)^2+\mathrm{i}c_n(\omega+j\omega_m)$, and $\hat{q}_n^{(j)}=\hat{k}_n^{(0)}+\mathrm{i}c_n(\omega+j\omega_m)$.

The vertical force at the base of the resonator can thus be obtained by substituting Eq. \eqref{equ:resonator disp FT} into Eq. \eqref{equ:resonator force}: 

\begin{equation} \label{equ:resonator force FT}
    F_n(t)=-m_n \frac{\partial^2 }{\partial t^2}\sum_{h=-\infty}^{\infty} \hat{W}_n^{(h)}\mathrm{e}^{\mathrm{i} (\omega+h\omega_m) t}=m_n \sum_{h=-\infty}^{\infty} (\omega+h\omega_m)^2\hat{W}_n^{(h)}\mathrm{e}^{\mathrm{i} (\omega+h\omega_m) t}=\sum_{h=-\infty}^{\infty} \hat{F}_n^{(h)}\mathrm{e}^{\mathrm{i} (\omega+h\omega_m) t},\quad h\in \mathbb{Z},
\end{equation}

\noindent where the $\hat{F}_n^{(h)}$  coefficients from $h=-P$ to $h=P$, collected in the vector $\mathbf{\hat{F}}_n$, read:

\begin{equation} \label{equ:force}
    \mathbf{\hat{F}}_n = \mathbf{D}_n \mathbf{\hat{W}}_n = \mathbf{D}_n \mathbf{M}_n^{-1} \mathbf{Q}_n \mathbf{\hat{w}}_n=:\mathbf{Z}_n\mathbf{\hat{w}}_n,
\end{equation}
with:

\begin{equation}
    \mathbf{\hat{F}}_n=\left[\begin{array}{c}
    \hat{F}_n^{(-P)} \\
    \hat{F}_n^{(-P+1)} \\
    \vdots \\
    \hat{F}_n^{(P)}
    \end{array}\right],\;
    \mathbf{D}_n=\left[\begin{array}{cccc}
    m_n(\omega-P\omega_m)^2 & 0 & \cdots & 0 \\
    0 & m_n[\omega+(-P+1)\omega_m]^2 & \cdots & 0 \\
    \vdots & \vdots & \ddots & \vdots \\
    0 & 0 & \cdots & m_n(\omega+P\omega_m)^2
    \end{array}\right].
\end{equation}

In Eq. \eqref{equ:force}, the matrix $\mathbf{Z}_n$ is the impedance operator which relates the resonator base motion to the resonator base force. It can be observed that the force exerted by each modulated resonator on the elastic substrate comprises multiple harmonics $(\omega+ h\omega_m)$. In the next section, we discuss how these forces generate the related multiple scattered wavefields.

\subsection{Elastic wave field of a finite cluster of modulated resonators}

We now consider an arbitrary distribution of $N$ space-time modulated resonators arranged on top of a given elastic waveguide. We assume that the resonators have an identical stiffness modulation frequency $\omega_m$. When an incident wave field $\mathbf{u}_0=[u_0,v_0,w_0]$ impinges the bases of such resonators, it triggers their vibrations which, in turn, generate scattered waves in the waveguide. Following a standard multiple scattering description \cite{torrent2013elastic,packo2019inverse,pu2022multiple}, the total wave field $\mathbf{u}=(u,v,w)$ at the generic position $\mathbf{r}$ along the waveguide can be expressed as the summation of the incident and scattered wave fields of the $N$ resonators:

\begin{subequations}
\begin{equation} \label{equ:total u}
     u(\mathbf{r},t)=u_0(\mathbf{r},t)+\sum_{n=1}^{N} F_n(t) G_u(\mathbf{r}-\mathbf{r}_n),
\end{equation}
\begin{equation} \label{equ:total v}
     v(\mathbf{r},t)=v_0(\mathbf{r},t)+\sum_{n=1}^{N} F_n(t) G_v(\mathbf{r}-\mathbf{r}_n),
\end{equation}
\begin{equation} \label{equ:total w}
     w(\mathbf{r},t)=w_0(\mathbf{r},t)+\sum_{n=1}^{N} F_n(t) G_w(\mathbf{r}-\mathbf{r}_n),
\end{equation}
\end{subequations}

\noindent where $G_u, G_v, G_w$ are the related Green's functions in terms of displacements along $x, y, z$. As in the previous section, we express the displacements of Eqs. \eqref{equ:total u}, \eqref{equ:total v}, \eqref{equ:total w} accounting for the multiple harmonics:

\begin{subequations}
\begin{equation} \label{equ:u FT}
    \sum_{h=-\infty}^{\infty} \hat{u}^{(h)}(\mathbf{r})\mathrm{e}^{\mathrm{i} (\omega+h\omega_m) t}=\sum_{h=-\infty}^{\infty} \hat{u}_0^{(h)}(\mathbf{r})\mathrm{e}^{\mathrm{i} (\omega+h\omega_m) t}+ \sum_{n=1}^{N}\sum_{h=-\infty}^{\infty} \hat{F}_n^{(h)}\hat{G}_u^{(h)}(\mathbf{r}-\mathbf{r}_n,\omega+h\omega_m)\mathrm{e}^{\mathrm{i} (\omega+h\omega_m) t}, \quad h\in \mathbb{Z}. 
\end{equation}

\begin{equation} \label{equ:v FT}
    \sum_{h=-\infty}^{\infty} \hat{v}^{(h)}(\mathbf{r})\mathrm{e}^{\mathrm{i} (\omega+h\omega_m) t}=\sum_{h=-\infty}^{\infty} \hat{v}_0^{(h)}(\mathbf{r})\mathrm{e}^{\mathrm{i} (\omega+h\omega_m) t}+ \sum_{n=1}^{N}\sum_{h=-\infty}^{\infty} \hat{F}_n^{(h)}\hat{G}_v^{(h)}(\mathbf{r}-\mathbf{r}_n,\omega+h\omega_m)\mathrm{e}^{\mathrm{i} (\omega+h\omega_m) t}, \quad h\in \mathbb{Z}. 
\end{equation}

\begin{equation} \label{equ:w FT}
    \sum_{h=-\infty}^{\infty} \hat{w}^{(h)}(\mathbf{r})\mathrm{e}^{\mathrm{i} (\omega+h\omega_m) t}=\sum_{h=-\infty}^{\infty} \hat{w}_0^{(h)}(\mathbf{r})\mathrm{e}^{\mathrm{i} (\omega+h\omega_m) t}+ \sum_{n=1}^{N}\sum_{h=-\infty}^{\infty} \hat{F}_n^{(h)}\hat{G}_w^{(h)}(\mathbf{r}-\mathbf{r}_n,\omega+h\omega_m)\mathrm{e}^{\mathrm{i} (\omega+h\omega_m) t}, \quad h\in \mathbb{Z}. 
\end{equation}
\end{subequations}

Truncating the harmonic terms from $h=-P$ to $h=P$, Eqs. \eqref{equ:u FT}, \eqref{equ:v FT}, \eqref{equ:w FT} can be rewritten as:

\begin{subequations}
\begin{equation} \label{equ:coeff of u}
    \mathbf{\hat{u}}(\mathbf{r})=\mathbf{\hat{u}}_0(\mathbf{r})+\sum_{n=1}^{N} \mathbf{\hat{G}}_u(\mathbf{r}-\mathbf{r}_n) \mathbf{\hat{F}}_n,
\end{equation}

\begin{equation} \label{equ:coeff of v}
    \mathbf{\hat{v}}(\mathbf{r})=\mathbf{\hat{v}}_0(\mathbf{r})+\sum_{n=1}^{N} \mathbf{\hat{G}}_v(\mathbf{r}-\mathbf{r}_n) \mathbf{\hat{F}}_n,
\end{equation}

\begin{equation} \label{equ:coeff of w}
    \mathbf{\hat{w}}(\mathbf{r})=\mathbf{\hat{w}}_0(\mathbf{r})+\sum_{n=1}^{N} \mathbf{\hat{G}}_w(\mathbf{r}-\mathbf{r}_n) \mathbf{\hat{F}}_n,
\end{equation}
\end{subequations}

\noindent with:


\begin{align*} 
    &\boldsymbol{\hat{\varphi}}(\mathbf{r})=\left[\begin{array}{c}
    \hat{\varphi}^{(-P)}(\mathbf{r}) \\
    \hat{\varphi}^{(-P+1)}(\mathbf{r}) \\
    \vdots \\
    \hat{\varphi}^{(P)}(\mathbf{r})
    \end{array}\right], \; 
    \boldsymbol{\hat{\varphi}}_0(\mathbf{r})=\left[\begin{array}{c}
    \hat{\varphi}_0^{(-P)}(\mathbf{r}) \\
    \hat{\varphi}_0^{(-P+1)}(\mathbf{r}) \\
    \vdots \\
    \hat{\varphi}_0^{(P)}(\mathbf{r})
    \end{array}\right], \; \varphi=u,v,w. \\
    &\mathbf{\hat{G}}_\varphi(\mathbf{r}-\mathbf{r}_n)=\left[\begin{array}{cccc}
    \hat{G}_\varphi^{(-P)}(\mathbf{r}-\mathbf{r}_n,\omega-P\omega_m) & 0 & \cdots & 0 \\
    0 & \hat{G}_\varphi^{(-P+1)}(\mathbf{r}-\mathbf{r}_n,\omega-P\omega_m+\omega_m) & \cdots & 0 \\
    \vdots & \vdots & \ddots & \vdots \\
    0 & 0 & \cdots & \hat{G}_\varphi^{(P)}(\mathbf{r}-\mathbf{r}_n,\omega+P\omega_m)
    \end{array}\right],
\end{align*}

\noindent and where $\boldsymbol{\hat{\varphi}}_0$ has non zero components only for the incident field $\varphi_0=u_0,v_0,w_0$:
\begin{equation}   
    \hat{\varphi}_0^{(j)} = \left\{
    \begin{array}{rl}
    \varphi_0 & j = 0\\
    0 & j \neq 0
    \end{array}, \quad j \in [-P,-P+1,...,P] \right.
\end{equation}

Note that in Eqs. \eqref{equ:coeff of u}, \eqref{equ:coeff of v}, \eqref{equ:coeff of w} the total displacement components $\mathbf{\hat{u}},\mathbf{\hat{v}},\mathbf{\hat{w}}$ and the elastic force coefficients $\mathbf{\hat{F}}_n$ are unknown.
Nonetheless, following a classical multiple scattering approach, we can obtain the coefficients $\mathbf{\hat{F}}_n$ by setting boundary conditions at resonator bases. In particular, we substitute Eq. \eqref{equ:force} into Eq. \eqref{equ:coeff of w} and specify it at the resonator location $\mathbf{r}_m$:

\begin{equation} \label{equ:total w at xm}
	\mathbf{Z}_m^{-1} \mathbf{\hat{F}}_m = \mathbf{\hat{w}}_0(\mathbf{r}_m) +\sum_{n=1}^{N}  \mathbf{\hat{G}}_w(\mathbf{r}_m-\mathbf{r}_n) \mathbf{\hat{F}}_n, \quad n, m=1,...,N.
\end{equation}

 Eq. \eqref{equ:total w at xm} leads to a system of $m=N$ equations that we can recast in matrix form as:

\begin{equation}
    \mathbf{AX}=\mathbf{B}, \label{equ:Ax=b}
\end{equation}
\noindent
with:

\begin{equation}
    \mathbf{A} = \left[\begin{array}{cccc}
    {\mathbf{Z}_1^{-1}-\mathbf{\hat{G}}_w(\mathbf{0})} & {-\mathbf{\hat{G}}_w(\mathbf{r}_1-\mathbf{r}_2)} & {\cdots} & {-\mathbf{\hat{G}}_w(\mathbf{r}_1-\mathbf{r}_N)} \\
    {-\mathbf{\hat{G}}_w(\mathbf{r}_2-\mathbf{r}_1)} & {\mathbf{Z}_2^{-1}-\mathbf{\hat{G}}_w(\mathbf{0})} & {\cdots} & {-\mathbf{\hat{G}}_w(\mathbf{r}_2-\mathbf{r}_N)} \\
    {\vdots} & {\vdots} & {\ddots} & {\vdots} \\
    {-\mathbf{\hat{G}}_w(\mathbf{r}_N-\mathbf{r}_1)} & {-\mathbf{\hat{G}}_w(\mathbf{r}_N-\mathbf{r}_2)} & {\cdots} & {\mathbf{Z}_N^{-1}-\mathbf{\hat{G}}_w(\mathbf{0})} \\
    \end{array}\right], \; \mathbf{X}=\left[\begin{array}{c}
    {\mathbf{\hat{F}}_1} \\
    {\mathbf{\hat{F}}_2} \\
    \vdots \\
    {\mathbf{\hat{F}}_N} \\  
    \end{array}\right], \;
    \mathbf{B}=\left[\begin{array}{c}
    {\mathbf{\hat{w}}_0(\mathbf{r}_1)} \\
    {\mathbf{\hat{w}}_0(\mathbf{r}_2)} \\
    \vdots \\
    {\mathbf{\hat{w}}_0(\mathbf{r}_N)} \\  
    \end{array}\right].
\end{equation}

It follows that for a given incident wave field $\mathbf{\hat{w}}_0$, the vector $\mathbf{X}$ of the force amplitudes $\mathbf{\hat{F}}_n$ can be computed as $\mathbf{X}=\mathbf{A}^{-1}\mathbf{B}$, and the total wave field in the waveguide evaluated by using Eqs. (\ref{equ:total u}), (\ref{equ:total v}), (\ref{equ:total w}).

In addition, we will show in the following subsection that Eq. \eqref{equ:Ax=b}, under appropriate assumptions, allows to derive the dispersion relation of time-modulated waveguides.

\subsection{Dispersion relation}

\begin{figure}[htbp]
\centering
\includegraphics[width=0.7\textwidth]{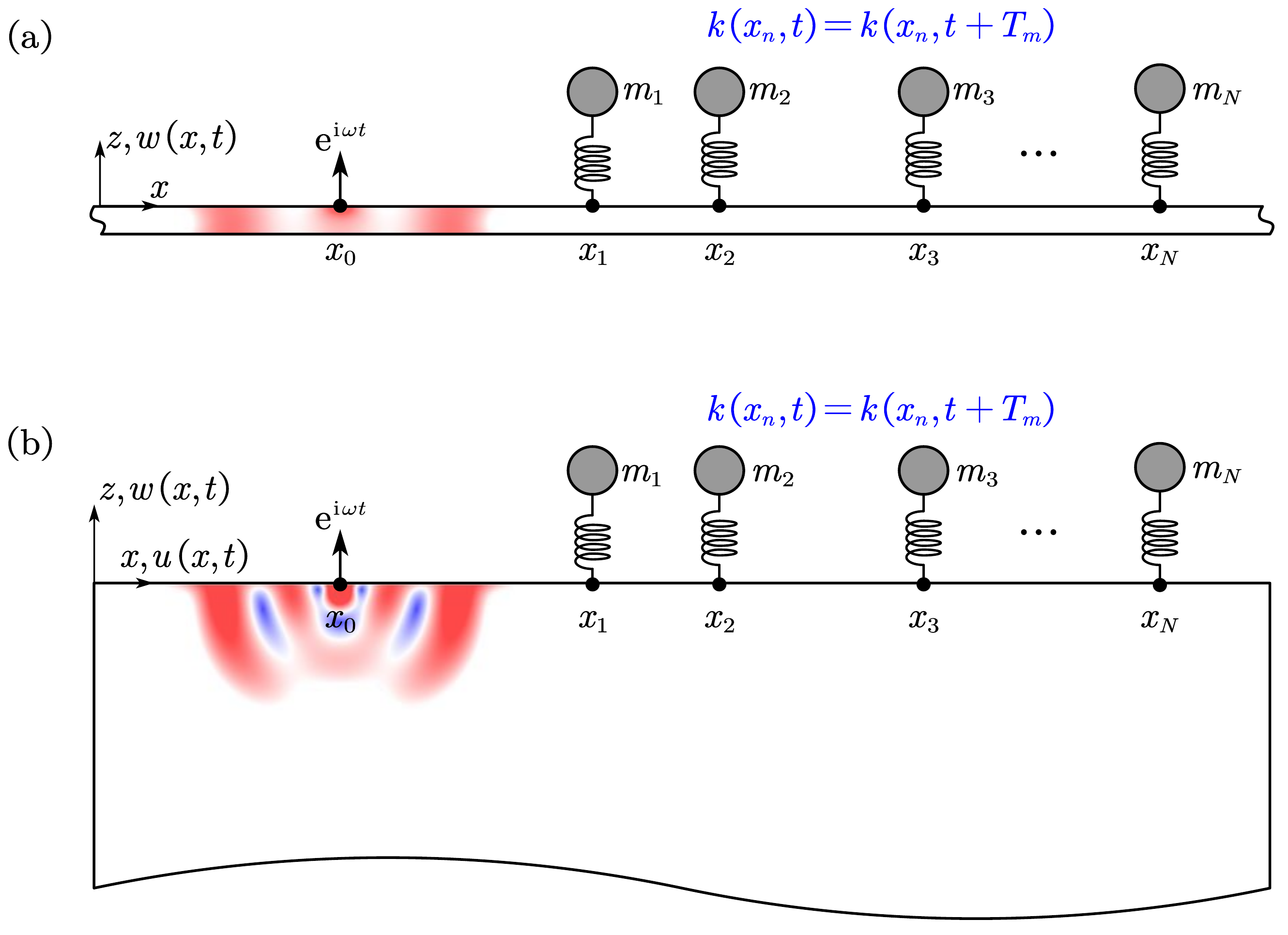}
\caption{Schematic of wave propagation in space-time modulated (a) metabeam and (b) metasurface. }
\label{fig:fig2}
\end{figure}

We here consider an infinite array of equally spaced resonators, arranged  atop an elastic waveguide (see Fig. \ref{fig:fig2}) at lattice distance $a$. We restrict our interest to oscillators with identical mass and with spring constant modulated in time and space with a wave-like modulation of period $T_m=2\pi/\omega_m$ and wavelength $\lambda_m=2\pi/\kappa_m$, whose general form reads:
\begin{equation}
k(x,t)=k(x+\lambda_m,t+T_m).
\end{equation}
\noindent
As before, we express $k(x,t)$ in a Fourier series form:
\begin{equation} \label{equ:FT K}
    k(x,t) = \sum_{j=-\infty}^{\infty} \hat{k}^{(j)}\mathrm{e}^{\mathrm{i} j(\omega_m t-\kappa_m x)}, \quad j\in \mathbb{Z}, 
\end{equation}
\noindent 
where the Fourier coefficients are computed as:

\begin{equation} \label{equ:FT coeff K}
    \hat{k}^{(j)}=\frac{\kappa_m}{2\pi} \frac{\omega_m}{2\pi} \int_{\frac{-\pi}{\kappa_m}}^{\frac{\pi}{\kappa_m}} \int_{\frac{-\pi}{\omega_m}}^{\frac{\pi}{\omega_m}}  k(x,t) \mathrm{e}^{-\mathrm{i} j(\omega_m t-\kappa_m x)}\,\mathrm{d}x\mathrm{d}t. 
\end{equation}
\noindent

As discussed in \cite{li2019transfer,palermo2020surface}, a stable response of the modulated system requires each modulation amplitude $\hat{k}^{(j)} (j \neq 0)$ to be sufficiently small with respect to the static stiffness $\hat{k}^{(0)}$. Under this assumption, for the assumed infinite ($N\rightarrow{\infty}$) periodic array of identical resonators, the scattering Eqs. \eqref{equ:total w at xm} are the same at any location $x_m$ and satisfy:

\begin{equation} \label{equ:disp at xm}
    \mathbf{Z}^{-1} \mathbf{\hat{F}}= \sum_{n=-N}^{N}  \mathbf{\hat{G}}_w(x_n) \mathbf{\hat{F}}(x_n)=\sum_{n=-\infty}^{\infty}  \mathbf{\hat{G}}_w(x_n) \mathbf{\hat{F}}(x_n)
\end{equation}


\noindent where $x_m$ has been conveniently set equal to 0.
\noindent
Following the effective medium approach \cite{garova1999interaction,boechler2013interaction,maznev2022effective}, namely considering the lattice spacing $a$ much smaller than the characteristic wavelength, we approximate the discrete point force as an average line load.
As a result, the total vertical displacement at the generic resonator base can be computed as:


\begin{equation} \label{equ:integral at xm}
    \mathbf{Z}^{-1} \mathbf{\hat{F}} = \frac{1}{a}\sum_{n=-\infty}^{\infty} \int_{x_n-a/2}^{x_n+a/2} \mathbf{\hat{G}}_w(x) \mathbf{\hat{F}}(x)\,\mathrm{d}x = \frac{1}{a} \int_{-\infty}^{\infty} \mathbf{\hat{G}}_w(x) \mathbf{\hat{F}}(x)\,\mathrm{d}x.
\end{equation}

Due to the space-time modulation of the resonator properties, we can express the force in Eq. \eqref{equ:resonator force FT} in the form:

\begin{equation} \label{equ:F(x,t)}
    F(x,t) = \sum_{h=-\infty}^{\infty} \hat{F}^{(h)}\mathrm{e}^{-\mathrm{i}(\kappa+h \kappa_m) x + \mathrm{i}(\omega+h \omega_m)t}, \quad h\in \mathbb{Z}. 
\end{equation}
\noindent
Substituting Eq. \eqref{equ:F(x,t)} into Eq. \eqref{equ:integral at xm} and truncating the orders from $h=-P$ to $h=P$ we obtain:


\begin{flalign}  \label{equ:infty integral}
    \nonumber
    & a\mathbf{Z}^{-1}\left[\begin{array}{c}
    \hat{F}^{(-P)} \\
    \hat{F}^{(-P+1)} \\
    \vdots \\
    \hat{F}^{(P)}
    \end{array}\right] = \\ 
    & \int_{-\infty}^{\infty} \left[\begin{array}{cccc}
    \hat{G}_w^{(-P)}(x,\omega-P\omega_m) & 0 & \cdots & 0 \\
    0 & \hat{G}_w^{(-P+1)}(x,\omega-P\omega_m+\omega_m) & \cdots & 0 \\
    \vdots & \vdots & \ddots & \vdots \\
    0 & 0 & \cdots & \hat{G}_w^{(P)}(x,\omega+P\omega_m)
    \end{array}\right] \left[\begin{array}{c}
    \hat{F}^{(-P)} \mathrm{e}^{-\mathrm{i}(\kappa-P \kappa_m) x}\\
    \hat{F}^{(-P+1)}\mathrm{e}^{-\mathrm{i}(\kappa-P \kappa_m+\kappa_m) x} \\
    \vdots \\
    \hat{F}^{(P)}\mathrm{e}^{-\mathrm{i}(\kappa+P \kappa_m) x}
    \end{array}\right]\,\mathrm{d}x.
\end{flalign} 

\noindent
Some minor algebra yields the following system of homogeneous equations:

\begin{small}
\begin{equation} \label{equ:FT integral}
    \left(a\mathbf{Z}^{-1}-
    \left[\begin{array}{cccc}
    \mathbf{\tilde{G}}(\kappa-P\kappa_m,\omega-P\omega_m) & 0 & \cdots & 0 \\
    0 & \mathbf{\tilde{G}}(\kappa-P\kappa_m+\kappa_m,\omega-P\omega_m+\omega_m) & \cdots & 0 \\
    \vdots & \vdots & \ddots & \vdots \\
    0 & 0 & \cdots & \mathbf{\tilde{G}}(\kappa+P\kappa_m,\omega+P\omega_m)
    \end{array}\right] \right) \left[\begin{array}{c}
    \hat{F}^{(-P)}\\
    \hat{F}^{(-P+1)}\\
    \vdots \\
    \hat{F}^{(P)}
    \end{array}\right]=\mathbf{0},
\end{equation}
\end{small}

\noindent
in which $\mathbf{\tilde{G}}(\kappa+P\kappa_m,\omega+P\omega_m)$ is $P$th order Green’s function in space-domain which is obtained as:

\begin{equation}
     \int_{-\infty}^{\infty} \hat{G}_w^{(P)}(x,\omega+P\omega_m) \mathrm{e}^{-\mathrm{i}(\kappa+P \kappa_m) x} \,\mathrm{d}x = \mathbf{\tilde{G}}(\kappa+P\kappa_m,\omega+P\omega_m).
\end{equation}
\noindent
Non-trivial solutions of Eq. \eqref{equ:FT integral} provide the dispersion relation:

\begin{small}
\begin{equation} \label{equ:dispersion equation}
    \mathcal{\tilde{C}}(\kappa, \omega) := \left |a\mathbf{Z}^{-1}-
    \left[\begin{array}{cccc}
    \mathbf{\tilde{G}}(\kappa-P\kappa_m,\omega-P\omega_m) & 0 & \cdots & 0 \\
    0 & \mathbf{\tilde{G}}(\kappa-P\kappa_m+\kappa_m,\omega-P\omega_m+\omega_m) & \cdots & 0 \\
    \vdots & \vdots & \ddots & \vdots \\
    0 & 0 & \cdots & \mathbf{\tilde{G}}(\kappa+P\kappa_m,\omega+P\omega_m)
    \end{array}\right] \right|=0.
\end{equation}
\end{small}

\section{Examples and applications}
\label{Examples and applications}

 To show the potential of our formulation, we consider two space-time-modulated waveguides that have been thoroughly discussed in previous works \cite{nassar2017EML,chen2019nonreciprocal,wu2021non,palermo2020surface}. We begin our investigation by considering an Euler beam coupled with an array of modulated resonators. For this example, we validate our approach against the results of Transfer Matrix Method (TMM). For the sake of completeness we report in \ref{Appendix A} the full derivation of the adopted TMM \cite{chen2019nonreciprocal}. As a second example, we consider a 2D elastic half-space coupled to modulated resonators. For this configuration, given the absence of closed-form formulations, we compare our findings vs. those obtained via finite element simulations, as in Ref. \cite{palermo2020surface}.

\subsection{Modeling non-reciprocal flexural waves in a space-time modulated beam}

We consider an Euler-Bernoulli beam equipped with an array of undamped resonators, see Fig. \ref{fig:fig2}a, modulated in a wave-like fashion according to the relationship \cite{nassar2017PRSA,nassar2017EML,chen2019nonreciprocal,vila2017bloch}:

\begin{equation} \label{equ:wave-like modulation}
    k(t,x)=k_{0}+k_a\cos(\omega_m t-\kappa_m x),
\end{equation}
\noindent where $k_{0}$ denotes the static stiffness, $k_a$ the amplitude of the modulation, $\omega_m$ the modulation angular frequency, $\kappa_m$ the modulation wavenumber. At any location $x_n$, the modulated stiffness is time-periodic and its non-zero Fourier coefficients read:


\begin{equation} 
    \hat{k}_n^{(0)}=k_{0},\quad \hat{k}_n^{(-1)}=\frac{1}{2}k_a\mathrm{e}^{\mathrm{i} \kappa_m x_n},\quad \hat{k}_n^{(1)}=\frac{1}{2}k_a\mathrm{e}^{-\mathrm{i} \kappa_m x_n},
\end{equation}
\noindent as $ \hat{k}_n^{(j)}=0$ for $|j|>1$.
For the numerical example, we consider the mechanical parameters originally adopted in Ref. \cite{nassar2017EML}, i.e., a resonator mass $m_0 = \rho \mathcal{A} a$, where $\rho$ is the mass density of the beam and $\mathcal{A}$ is the cross-section area of the beam. Similarly, the modulation frequency is set as $\omega_m=0.25\omega_0$ and modulation amplitude as $k_a = 0.2 k_0$, in which $\omega_0$ is the resonance frequency of resonators and $k_0=m_0 \omega_0^2$ is the unmodulated stiffness; the modulation wavenumber is $\kappa_m=1.25\kappa_0$, where $\kappa_0=\sqrt[4]{k_0/(a\mathcal{D})}$, $\mathcal{D}$ the bending stiffness of the beam.

\subsubsection{Dispersion relation}
According to the Euler–Bernoulli beam theory, the $P$th order governing equation under the action of a harmonic point force can be written as:

\begin{equation} \label{equ:Euler beam equation}
    \mathcal{D}\frac{\partial^4 w}{\partial x^4}+\rho \mathcal{A}\frac{\partial^2 w}{\partial t^2}=\delta \left( x \right) \mathrm{e}^{\mathrm{i} \omega_P t}, \quad P\in \mathbb{Z}.
\end{equation}
\noindent
We Fourier transform Eq. \eqref{equ:Euler beam equation} along the $x$ direction, and obtain the transformed $P$th order Green’s function in space-domain as:

\begin{equation} \label{equ:space-domain GF of Euler beam}
    \tilde{G}(\kappa_P,\omega_P)=\frac{1}{\mathcal{D} \kappa_P^4-\rho \mathcal{A} \omega_P^2},
\end{equation}
\noindent
where the shifted frequency and wavenumber are defined as:

\begin{equation} \label{equ:shifted frequency and wavenumber}
    \omega_P = \omega+P \omega_m, \quad \kappa_P = \kappa+P \kappa_m, \quad P\in \mathbb{Z}.
\end{equation}
\noindent

First, by setting $P=0$ we get the non-modulated impedance parameter $Z$ from Eq. \eqref{equ:force} as:

\begin{equation} \label{equ:impedance of non-modulated resonator}
    Z = \frac{m_0 \omega_0^2 \omega^2}{\omega_0^2-\omega^2}.
\end{equation}
\noindent
Substituting Eqs. \eqref{equ:space-domain GF of Euler beam} and \eqref{equ:impedance of non-modulated resonator} into Eq. \eqref{equ:dispersion equation} we obtain immediately the dispersion relation of a non-modulated metabeam:

\begin{equation} \label{equ:original dispersion equation of metabeam}
    \mathcal{C}(\kappa,\omega) := \mathcal{D} \kappa^4-\left(\rho \mathcal{A}+\frac{m_0}{a} \frac{1}{1-\omega^2 / \omega_0^2}\right) \omega^2 = 0.
\end{equation}

\noindent
This dispersion equation is identical to the one obtained in Refs. \cite{nassar2017EML,chen2019nonreciprocal} and matches the dispersion curve provided by FE simulations,  see \ref{Appendix B} for details.

\begin{figure}[htbp]
\centering
\includegraphics[width=1\textwidth]{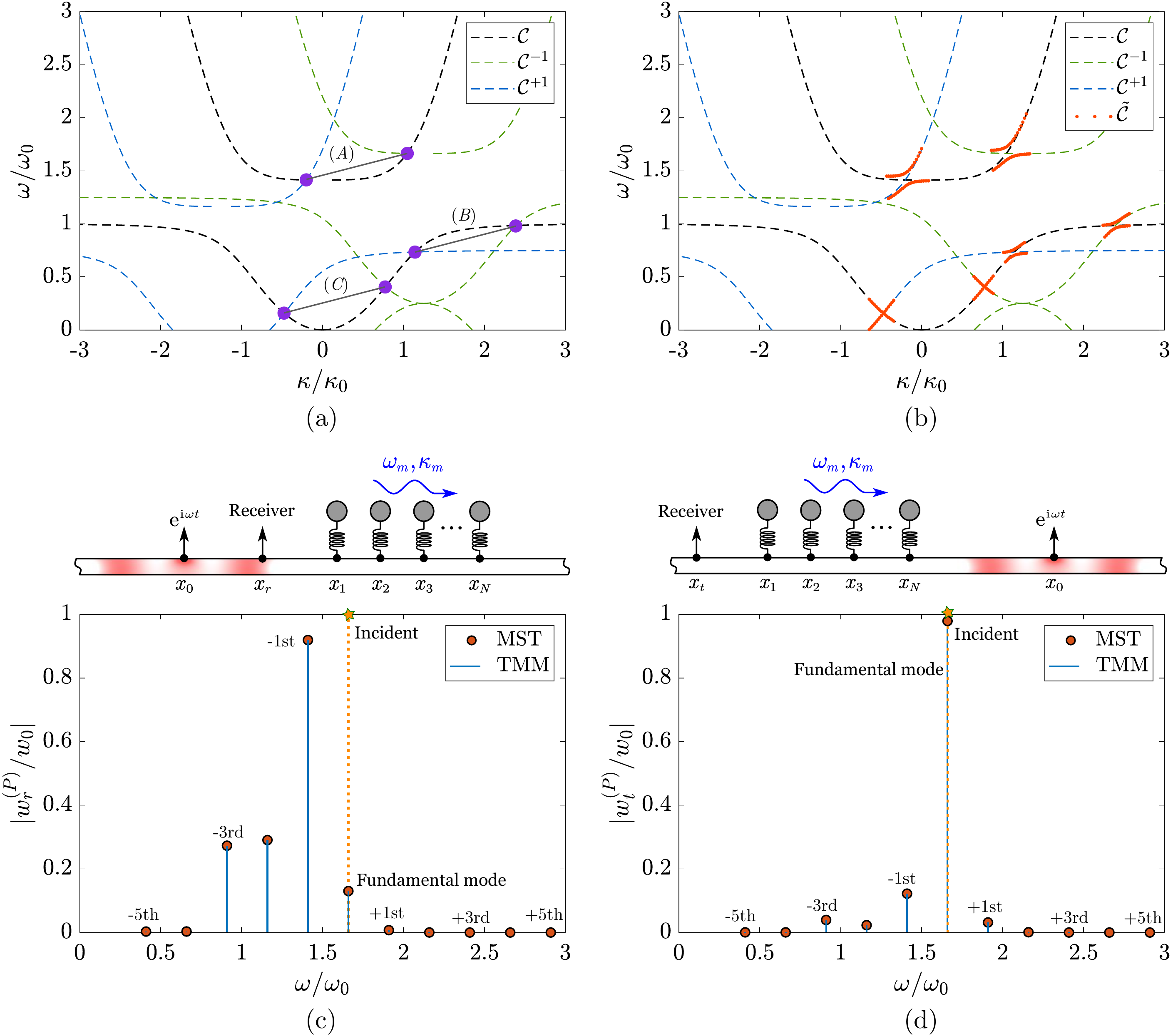}
\caption{(a) Dispersion curve of a non-modulated metabeam (black dashed lines) and its shifted analogs for $P=-1$ and $P=1$, respectively. (b) Dispersion curves (circular red markers) of a modulated metabeam in proximity of the phase matching pairs. Normalized reflection and transmission coefficients for flexural waves propagating at $\omega=1.66\omega_0$ inside the directional band gap (pair $A$) for (c) a right and (d) a left traveling incident wave (star marker), respectively. For comparison, results obtained by the transfer matrix method (TMM) are also provided (blue solid lines). 
}
\label{fig:fig3}
\end{figure}

In the presence of modulation, scattered waves are expected when the phase matching condition is satisfied, i.e., $\mathcal{C}(\kappa,\omega) = \mathcal{C}(\kappa_P, \omega_P)$ \cite{nassar2017PRSA}. As an example, in Fig. \ref{fig:fig3}a we show the original ($\mathcal{C}$) and the two shifted dispersion curves ($\mathcal{C}^{\pm 1}$) for $P=\pm 1$, respectively. The phase matching condition is met at the intersections between the original curve and the shifted ones, namely at six magenta points of the pairs $(A)$, $(B)$ and $(C)$. The asymmetric distribution of these intersections suggests the breaking of time-reversal symmetry which, in turn, leads to direction-dependent phenomena within the metabeam \cite{chen2019nonreciprocal}. 

We now predict the dispersion properties of the modulated metabeam. To do so, we substitute Eq. \eqref{equ:space-domain GF of Euler beam} into Eq. \eqref{equ:dispersion equation} by truncating waves to the first order ($P=1$), which yields:

\begin{equation} \label{equ:dispersion equation of Euler beam}
    \mathcal{\tilde{C}}(\kappa, \omega) := \left |a\mathbf{Z}^{-1}-
    \left[\begin{array}{ccc}
    1/[\mathcal{D} (\kappa-\kappa_m)^4-\rho \mathcal{A} (\omega-\omega_m)^2] & 0  & 0 \\
    0 & 1/[\mathcal{D} \kappa^4-\rho \mathcal{A} \omega^2]  & 0 \\
    0 & 0 & 1/[\mathcal{D} (\kappa+ \kappa_m)^4-\rho \mathcal{A} (\omega+ \omega_m)^2]
    \end{array}\right] \right|=0,
\end{equation}

\noindent
with the impedance operator:

\begin{small}
\begin{equation}
\begin{aligned}
&\mathbf{Z}=m_0\left[\begin{array}{rrr}
(\omega-\omega_m)^2 & 0 & 0  \\
0 & \omega^2 & 0 \\
0 & 0 & (\omega+ \omega_m)^2
\end{array}\right]
\left[\begin{array}{rrr}
k_0-m_0(\omega-\omega_m)^2 & 0.5k_a & 0  \\
0.5k_a & k_0-m_0\omega^2 & 0.5k_a  \\
0 & 0.5k_a & k_0-m_0(\omega+\omega_m)^2
\end{array}\right]^{-1}
\left[\begin{array}{rrr}
k_0 & 0.5k_a & 0  \\
0.5k_a & k_0 & 0.5k_a \\
0 & 0.5k_a & k_0
\end{array}\right]. 
\end{aligned}
\end{equation}
\end{small}

We remark that the coupled dispersion relation in Eq. \eqref{equ:dispersion equation of Euler beam} holds only near the above-mentioned intersections in Fig. \ref{fig:fig3}a \cite{wang2018observation}. Thus, we compute and plot the coupled dispersion in the range of $\pm 0.1\kappa$ and $\pm 0.1\omega$ around each crossing point, as shown in Fig. \ref{fig:fig3}b (red circular markers). For comparison, we also provide the unmodulated  dispersion curve (solution of Eq. \eqref{equ:original dispersion equation of metabeam}) and its shifted analogs on the same figure. As discussed in Ref. \cite{nassar2017EML}, in the vicinity of pair $B$ no directional band gap is generated, since both modes have positive group velocities. Conversely, for contra-directional branches such as pairs $A$ and $C$, the repulsion effect can lead to narrow band directional gaps, for instance, around $A$. Within these gaps, 
waves are hindered only when propagating along the specific direction (dictated by the sign of the related wavenumber); conversely, they are fully transmitted when propagating along the opposite direction \cite{nassar2017PRSA}. This directional wave-filtering is usually accompanied by the generation of lower/higher-order waves at the phase-matched frequencies, thus resulting in a reflection combined with a frequency conversion  \cite{nassar2017EML}. Evidence of these effects is provided in the next section where the steady-state solution of waves propagating along a finite modulated metabeam is computed.

\subsubsection{Steady-state solutions}

To evidence the non-reciprocal behavior predicted by the dispersion analysis, we utilize Eq. \eqref{equ:coeff of w} to compute the steady-state response of a finite metabeam. In particular, we are interested in verifying the non-reciprocal reflection/transmission in the directional band gap at pair $A$ in Fig. \ref{fig:fig3}. As an example, an array of 50 resonators is considered for these investigations. The response is recorded at locations $x_r$ and $x_t$, and later used to compute the reflection and transmission values, respectively. In both scenarios, the harmonic point source $\mathrm{e}^{\mathrm{i} \omega t}$ and the receiver are located at distances of $d_s=600a$ and $d_r=300a$ from the closest oscillator.

According to the formulation discussed in Section \ref{Theoretical formulation}, the impedance operators $\mathbf{Z}_1$ to $\mathbf{Z}_N$ are obtained from Eq. \eqref{equ:force} while the $P$th order Green's function in Eq. \eqref{equ:Ax=b} is obtained by applying the inverse Fourier transform to Eq. \eqref{equ:space-domain GF of Euler beam} as:

\begin{equation} \label{equ:Green's function of Euler beam}
    \hat{G}_w^{(P)}(x, \omega+P \omega_m) = \frac{-1}{4\mathcal{D}\beta_P^3}(\mathrm{e}^{-\beta_P|x|}+\mathrm{i}\,\mathrm{e}^{\mathrm{-i}\beta_P|x|}),
\end{equation}
\noindent
where the $P$th order wavenumber for flexural waves reads:

\begin{equation}
    \beta_P=\sqrt[4]{\frac{\rho \mathcal{A} (\omega+P \omega_m)^2}{\mathcal{D}}}.
\end{equation}

Substituting Eq. \eqref{equ:Green's function of Euler beam} into Eq. \eqref{equ:Ax=b} we obtain the elastic force coefficients $\mathbf{\hat{F}}_n$, which are inserted into Eq. \eqref{equ:coeff of w} for the calculation of the displacement components $\mathbf{\hat{w}}(x)$ in the beam.

We begin our investigation considering a right-propagating ($\kappa>0$) flexural wave at frequency $\omega=1.66\omega_0$, i.e., the intersection at pair $A$ in Fig. \ref{fig:fig3}a. The reflection  coefficient, normalized with respect to the incident wave, $|w_r/w_0|$, is displayed in Fig. \ref{fig:fig3}c, considering the scattered waves truncated at $P=\pm5$ order.

As expected, right-propagating incident waves at $\omega=1.66\omega_0$ (dashed line in Fig. \ref{fig:fig3}c) undergo a strong reflection with different frequency contents, including the largest component at the first-order harmonic ($1.66\omega_0-\omega_m$) and non-negligible components at the second ($1.66\omega_0-2\omega_m$) and third-order harmonic ($1.66\omega_0-3\omega_m$). The amplitude of other higher-order harmonics is negligible. Conversely, the left-propagating wave (opposite to the modulation direction) with the same frequency $\omega=1.66\omega_0$ can travel through the resonators almost undisturbed, as shown by the normalized transmission $|w_t/w_0|$ in Fig. \ref{fig:fig3}d. To verify the predictions provided by our approach, we compute the same transmission and reflection coefficients using the transfer matrix method. The results, marked by solid lines in Figs. \ref{fig:fig3}c,d, are in excellent agreement with our analytical solutions (see more details on the transfer matrix method in \ref{Appendix A}).

\subsection{Modeling non-reciprocal Rayleigh wave propagation in a space-time modulated metasurface}

We now consider the propagation of Rayleigh waves across a cluster of modulated resonators. Such a problem has been recently investigated with the aid of FE numerical simulations \cite{wu2021non, palermo2020surface}. Our purpose is to show the capability of the proposed analytical formulation to reproduce both the non-reciprocal dispersion and the reflection/transmission coefficients in this complex configuration. 

For our example, we consider the parameters recently used in Ref. \cite{palermo2020surface}: a half-space with $c_L/c_T=2$, a resonator with mass ratio $m_0 \omega_0/(\rho a c_T)=0.15$, the modulation frequency $\omega_m/\omega_0=0.25$, and the modulation wavenumber $\kappa_m/\kappa_r=2.5$, in which $\kappa_r=\omega_0/c_T$. 

\subsubsection{Dispersion relation}
Let us briefly recall the Green's function  for a 2D isotropic elastic half-space actuated by a harmonic vertical load acting at the surface. For this configuration, the equilibrium equation  can be formulated as a boundary value problem:

\begin{subequations}
\begin{equation} \label{equ:Navier's Equation}
    c_L^2 \nabla(\nabla \cdot \mathbf{u})-c_T^2 \nabla \times(\nabla \times \mathbf{u})= \frac{\partial^2 \mathbf{u}}{\partial t^2}, \quad \text{for} \;z<0,
\end{equation}

\begin{equation} \label{equ:BC of Navier's Equation}
    \tau_{zx}(x,0) = 0, \quad \sigma_z(x,0)=\delta(x) \mathrm{e}^{\mathrm{i} \omega_P t}, 
\end{equation}
\end{subequations}
\noindent
in which $c_L$ and $c_T$ denote the longitudinal (L) and transverse (T) wave velocities, and $\tau_{zx}$, $\sigma_z$ represent the shear and normal stresses, respectively; $\mathbf{u}$ is the displacement field with components $u$ and $w$; $\delta(x)$ is the Dirac delta function.

In analogy with the metabeam problem, we Fourier transform the equilibrium  Eqs. \eqref{equ:Navier's Equation} and \eqref{equ:BC of Navier's Equation} along the $x$ direction, and obtain the transformed $P$th order Green’s function at $z=0$ as:

\begin{equation} \label{equ:space-domain GF of Lamb's problem}
    \tilde{G}(\kappa_P,\omega_P)=\frac{1}{\rho c_T^4} \frac{\omega_P^2\sqrt{\kappa_P^2-\frac{\omega_P^2}{c_L^2}}}{4 \kappa_P^2\sqrt{\left(\kappa_P^2-\frac{\omega_P^2}{c_T^2}\right)\left(\kappa_P^2-\frac{\omega_P^2}{c_L^2}\right)}-\left(2 \kappa_P^2-\frac{\omega_P^2}{c_T^2}\right)^2},
\end{equation}
\noindent
where $\rho$ is the density of the substrate, and the shifted frequency $\omega_P$ and wavenumber $\kappa_P$ are defined in Eq. \eqref{equ:shifted frequency and wavenumber}.
Substituting Eqs. \eqref{equ:impedance of non-modulated resonator} and \eqref{equ:space-domain GF of Lamb's problem} into Eq. \eqref{equ:dispersion equation} and setting $P=0$, we obtain immediately the dispersion relation for Rayleigh waves existing in a non-modulated metasurface:

\begin{equation} \label{equ:original dispersion equation of metasurface}
    \mathcal{C}(\kappa,\omega) := \left(2 \kappa^2-\frac{\omega^2}{c_T^2}\right)^2 - 4 \kappa^2\sqrt{\left(\kappa^2-\frac{\omega^2}{c_T^2}\right)\left(\kappa^2-\frac{\omega^2}{c_L^2}\right)}- \frac{m_0 \omega^4 \sqrt{\kappa^2-\frac{\omega^2}{c_L^2}}}{\rho a c_T^4 (\omega^2/\omega_0^2 -1)}=0.
\end{equation}
\noindent
This dispersion equation is identical to the one obtained in Refs. \cite{boechler2013interaction,maznev2022effective}, and matches the numerical dispersion curve computed via  FEM, see \ref{Appendix B}.

As for the metabeam scenario, we first plot the unmodulated $\mathcal{C}(\kappa,\omega)$ and the shifted $\mathcal{C}(\kappa_P, \omega_P)$ dispersion curves for $P=\pm 1$, Fig. \ref{fig:fig4}a. Again, phase matching points (e.g., pairs $A$ to $E$) are found when $\mathcal{C}(\kappa,\omega) = \mathcal{C}(\kappa_P, \omega_P)$ is met. We predict the dispersion properties of the modulated metasurface around these points using Eq. \eqref{equ:dispersion equation}. To this purpose, we substitute Eq. \eqref{equ:space-domain GF of Lamb's problem} into Eq. \eqref{equ:dispersion equation} and truncate the expansion to the first order, using the impedance operator $\mathbf{Z}$ computed according to Eq. \eqref{equ:original dispersion equation of metabeam}.

We display the modulated dispersion relation in the range of $\pm 0.1\kappa$ and $\pm 0.1\omega$ around each intersection in Fig. \ref{fig:fig4}b (red circular markers). As an example, the intersection between contra-directional branches gives rise to the locking pair $C$ which results in a directional band gap. Harmonic waves propagating with wavenumber-frequency falling within the directional gap ($1.21\kappa_r, 1.185\omega_0$) are reflected by the metasurface as a propagating mode at the phase-matched frequency-wavenumber pair ($1.21\kappa_r-\kappa_m, 1.185\omega_0-\omega_m$). Conversely, such reflection by conversion does not occur for waves propagating along the opposite direction at the same frequency $1.185\omega_0$, confirming the non-reciprocity due to the broken time-reversal symmetry \cite{palermo2020surface}. Again, clear evidence of these effects predicted by the dispersion curve is provided in the next section by computing the steady-state solutions of Rayleigh waves propagating along a finite modulated metasurface.

\subsubsection{Steady-state solutions}
\label{Steady-state solutions of SAW}

To shown the non-reciprocal Rayleigh wave propagation discussed above, we use Eq. \eqref{equ:coeff of w} to compute the steady-state response of a finite metasurface. The impedance operators $\mathbf{Z}_1$ to $\mathbf{Z}_N$ are computed from Eq. \eqref{equ:force}, while the Green's function in Eq. \eqref{equ:Ax=b} is obtained by applying the inverse Fourier transform to Eq. \eqref{equ:space-domain GF of Lamb's problem}, yielding the $P$th order wave field (Green's function) at $z=0$:

\begin{equation} \label{equ:Green's function of Lamb's problem}
    \hat{G}_w^{(P)}(x,0,\omega_P) = \frac{1}{2 \pi \rho c_T^4} \int_{-\infty}^{\infty}  \frac{\omega_P^2\sqrt{\kappa_P^2-\frac{\omega_P^2}{c_L^2}}}{4 \kappa_P^2\sqrt{\left(\kappa_P^2-\frac{\omega_P^2}{c_T^2}\right)\left(\kappa_P^2-\frac{\omega_P^2}{c_L^2}\right)}-\left(2 \kappa_P^2-\frac{\omega_P^2}{c_T^2}\right)^2} \mathrm{e}^{\mathrm{i} \kappa_P x} \mathrm{~d} \kappa_P.
\end{equation}

We note that unlike the Green's function of an Euler beam, Eq. \eqref{equ:Green's function of Euler beam}, the one for a 2D elastic substrate is divergent at the origin, Eq. \eqref{equ:Green's function of Lamb's problem}. To avoid any convergence issue, we introduce a small footprint of length $\ell_s$ for each resonator so that the associated Green's functions are \cite{pu2022multiple}:

\begin{subequations}
\begin{equation} \label{equ:G_w}
     \hat{G}_w^{(P)}(x,z,\omega_P)=\frac{1}{\pi \rho c_T^2} \int_{-\infty}^{\infty}\frac{\sin(\kappa_P\ell_s/2)}{\kappa_P}\frac{2\kappa_P^2\beta_L\mathrm{e}^{\beta_Tz}-\beta_L(2\kappa_P^2-\frac{\omega_P^2}{c_T^2})\mathrm{e}^{\beta_Lz}}{4 \kappa_P^2 \beta_L \beta_T-\left(2 \kappa_P^2-\frac{\omega_P^2}{c_T^2}\right)^2} \mathrm{e}^{\mathrm{i}\kappa_Px}\,\mathrm{d}\kappa_P,
\end{equation}
for the vertical displacement components and
\begin{equation} \label{equ:G_u}
     \hat{G}_u^{(P)}(x,z,\omega_P)=\frac{\mathrm{i}}{\pi \rho c_T^2} \int_{-\infty}^{\infty}\sin(\kappa_P\ell_s/2) \frac{2\beta_L \beta_T\mathrm{e}^{\beta_Tz}-(2\kappa_P^2-\frac{\omega_P^2}{c_T^2})\mathrm{e}^{\beta_Lz}}{4 \kappa_P^2 \beta_L \beta_T-\left(2 \kappa_P^2-\frac{\omega_P^2}{c_T^2}\right)^2} \mathrm{e}^{\mathrm{i}\kappa_Px}\,\mathrm{d}\kappa_P, 
\end{equation}
\end{subequations}
for the horizontal ones, where: 

\begin{equation}
    \beta_L = \sqrt{\kappa_P^2-\frac{\omega_P^2}{c_L^2}}, \quad \beta_T = \sqrt{\kappa_P^2-\frac{\omega_P^2}{c_T^2}}.
\end{equation}

Substituting Eq. \eqref{equ:G_w} into Eq. \eqref{equ:Ax=b} we obtain the elastic force coefficients $\mathbf{\hat{F}}_n$, which are used in Eqs. \eqref{equ:coeff of u}, \eqref{equ:coeff of w} to compute the wave field $\mathbf{\hat{u}}(x,z)$, $\mathbf{\hat{w}}(x,z)$ in the substrate.

\begin{figure}[htbp]
\centering
\includegraphics[width=1\textwidth]{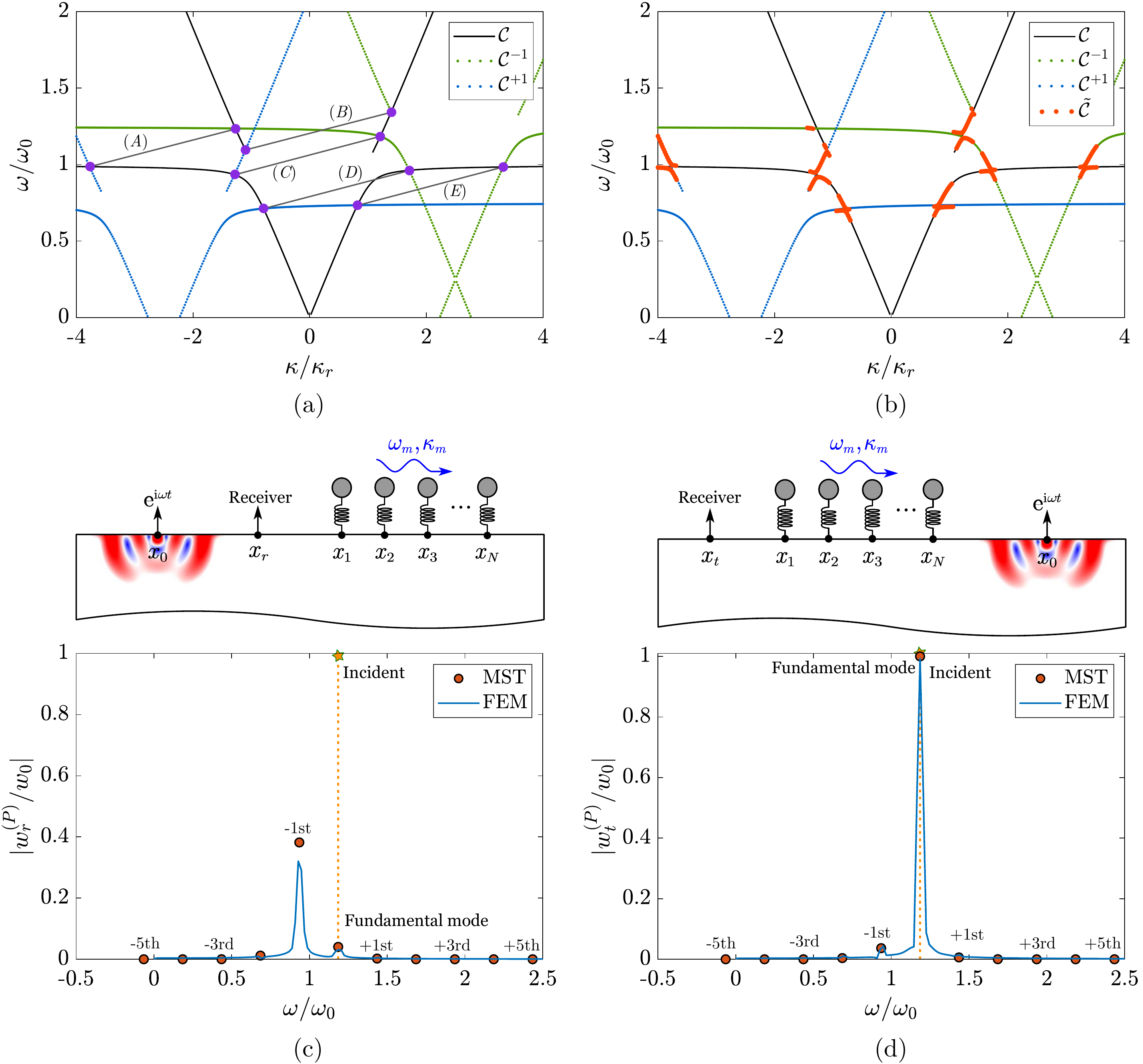}
\caption{Analytical modeling of a metasurface. (a) Dispersion relation of a non-modulated metasurface (black solid lines) and its shifted analogs for $P=-1$ and $P=1$, respectively. (b) Dispersion relation (circular markers) of a modulated metasurface in the vicinity of phase matching pairs. Steady-state solutions of Rayleigh wave propagation at $\omega=1.185\omega_0$ inside the narrow directional band gap (pair $C$) for (c) a right and (d) a left traveling incident wave (star marker), respectively.}
\label{fig:fig4}
\end{figure}

For our example, we compute the steady-state response at locations $(x_r,0)$ and $(x_t,0)$ on the substrate surface considering an array of 100 resonators with footprint width $\ell_s=a/20$, where the harmonic point source and the receiver are located at distances of $d_s=600a$ and $d_r=300a$ from the closest resonator. For a right-going ($\kappa>0$) incident Rayleigh wave (dashed line) at $\omega=1.185\omega_0$ the reflected field, shown in Fig. \ref{fig:fig4}c, confirms a back-scattering at the coupled frequency $\omega=1.185\omega_0-\omega_m$. Conversely, the left-going ($\kappa<0$) incident Rayleigh wave (dashed line) at the same frequency $\omega=1.185\omega_0$ propagates without  reflection or frequency conversion phenomena (see Fig. \ref{fig:fig4}d).

We now resort to the FEM to verify our analytical solutions. To this purpose, we build a 2D plane-strain model in a commercial FE software (COMSOL Multiphysics). The reader can find the details of the numerical model in \ref{Appendix C}. Specifically, we compute the transient response of the system actuated by a vertical tone-burst-shaped force having central frequency $\omega=1.185\omega_0$, and analyze the vertical displacement field $w$ (see the insets of Fig. \ref{fig:fig4}). The corresponding frequency spectra (solid line) computed through the Fourier transform (FFT) of the record time-domain data at the receiver are displayed in Figs. \ref{fig:fig4}c,d.
The reader can appreciate how the numerical results match the analytical solutions.

\begin{figure}[htbp]
\centering
\includegraphics[width=1\textwidth]{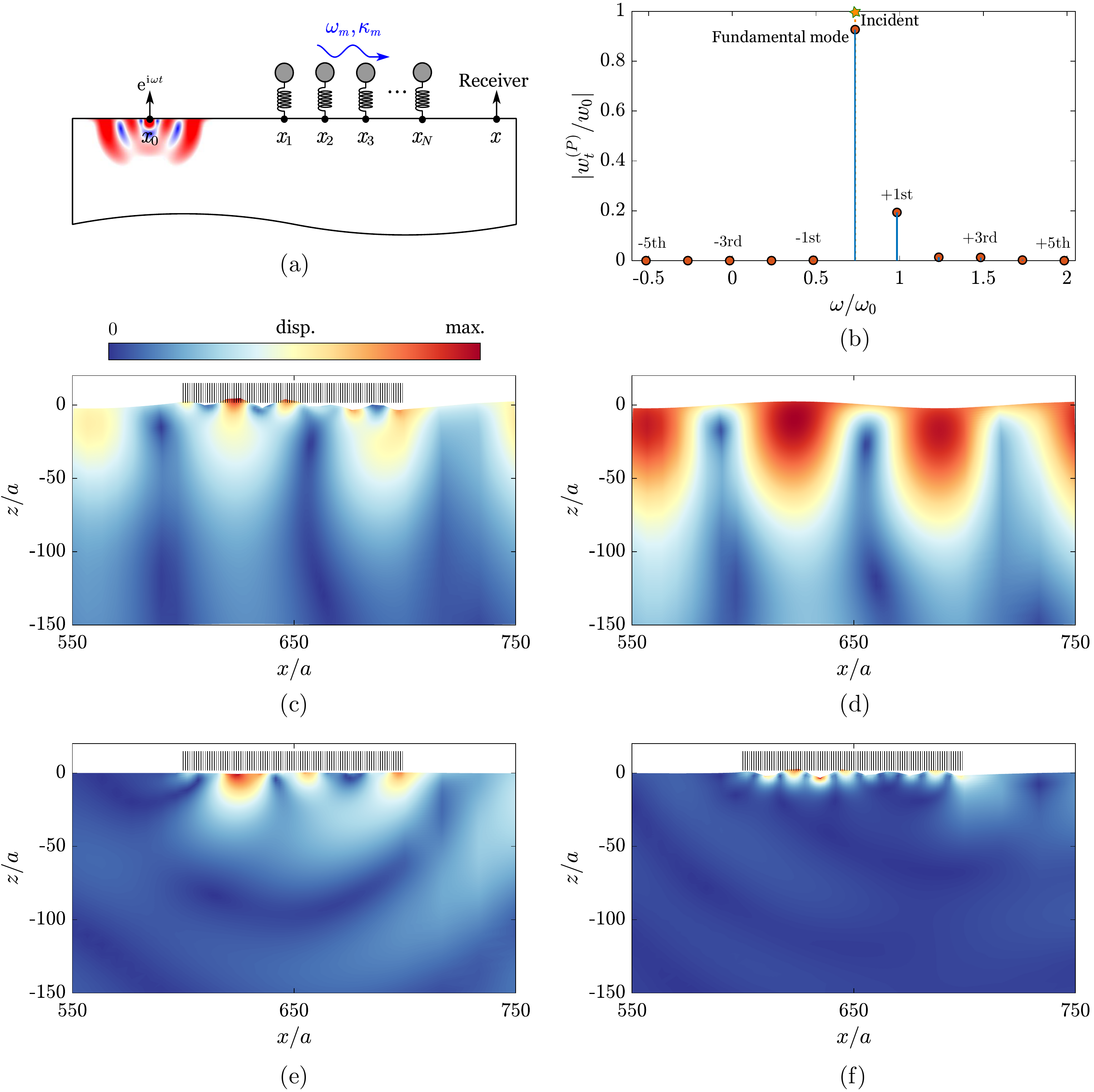}
\caption{Harmonic responses of a modulated metasurface. (a) Schematic for right-propagating Rayleigh waves. (b) Steady-state solutions of Rayleigh wave propagation at $\omega=0.734\omega_0$ (pair $E$ in Fig. \ref{fig:fig4}). The total wave field, free field, fundamental scattered field ($\omega=0.734\omega_0$), and the first-order scattered field ($\omega=0.734\omega_0+\omega_m$) excited by the source at $\omega=0.734\omega_0$ are shown in (c), (d), (e) and (f), respectively.}
\label{fig:fig5}
\end{figure}

Finally, we inspect the steady-state response at the ``veering pair" (intersection between two co-directional branches) \cite{palermo2020surface} $E$ in Fig. \ref{fig:fig4}, where we expect the Rayleigh wave to be transmitted and converted from one harmonic to another \cite{wu2021non,palermo2020surface}. To evidence such a conversion, we utilize the same model (see Fig. \ref{fig:fig5}) excited by a right-going incident Rayleigh wave at frequency $\omega=0.734\omega_0$. The results obtained with both the analytical solutions and the FE model are collected in Fig. \ref{fig:fig5}b. As expected, the modulated metasurface can convert the incident wave ($\omega=0.734\omega_0$) into a transmitted wave with a different frequency content, e.g., the phase matched first-order harmonic at $\omega=0.734\omega_0+\omega_m$.

To better appreciate this effect, we compute the total wave field $\sqrt{\Re{(u)}^2+\Re{(w)}^2}$,  using Eqs. \eqref{equ:coeff of u}, \eqref{equ:coeff of w}, \eqref{equ:G_w}, \eqref{equ:G_u}, in the domain $x=[550,750]a$, $z=[-150,0]a$. The total wave field, shown in Fig. \ref{fig:fig5}c, can be decomposed by Eqs. \eqref{equ:coeff of u}, \eqref{equ:coeff of w} into the incident field at $\omega=0.734\omega_0$, Fig. \ref{fig:fig5}d,  and scattered wave fields: the fundamental mode at $\omega=0.734\omega_0$ in Fig. \ref{fig:fig5}e and the first-order harmonic at $\omega=0.734\omega_0+\omega_m$ in Fig. \ref{fig:fig5}f. Both scattering fields exhibit a clear asymmetry, with the right-hand side having a greater amplitude than the left-hand side, a clear feature of the forward scattering behavior at veering pairs of the modulated metasurface.

\section{Conclusion} \label{Conclusion}

We developed a multiple scattering formulation to model the interaction of a given incident field with a cluster of space-time-modulated resonators located at the surface of a given elastic waveguide. The effect of time-varying resonators is modeled by means of impedance operators, able to account for lower- and higher-order harmonics generated by the modulated oscillators. The vertical motion of resonators, actuated by the incident field, generates scattered fields in the waveguide, which are characterized via ad-hoc Green’s functions. The unknown amplitudes of scattered fields are then obtained from a multiple scattering scheme by ensuring the continuity of displacement at the footprint of resonators. 

We have demonstrated the capabilities and accuracy of our framework by computing both the dispersion relation and wave field of flexural and Rayleigh waves propagating along modulated beams and substrates, respectively.

Our approach has several advantages compared to currently available methods for studying elastic waves along space-time-modulated metamaterials. First, it allows to investigate an arbitrary number of resonators with no restriction on their spatial configuration and modulation profile, apart from their common modulation period $T_m$. Second, it enables the analytical treatment of non-reciprocal wave propagation in higher dimensional systems (2D and 3D), thus overcoming the limitation of currently available analytical methods (e.g, the transfer matrix method) valid only for 1D wave propagation problems \cite{li2019transfer}. Third, our method is able to reduce the computational cost with respect to classical numerical schemes since it does not require the discretization of the entire system. This feature is particularly appealing for modeling wave propagation in higher dimensional systems and will prove its value for future design and optimization studies. Fourth, it advances the knowledge of multiple scattering theory which has demonstrated its superior capabilities in modeling the interaction of oscillators with elastic flexural and surface acoustic waves \cite{torrent2013elastic,packo2019inverse,pu2022multiple,PU2022topological}. 
 
Overall, we anticipate the proposed formulation will serve as a powerful tool to explore various modulation profiles on elastic waveguides and to guide future experiments on space-time-modulated systems. Since the framework developed in this work is general, we also expect an extension into acoustics and electromagnetism, thus supporting the development of  nonreciprocal devices for both acoustics and optics.

\section*{CRediT authorship contribution statement}
\noindent \textbf{Xingbo Pu:} Conceptualization, Methodology, Investigation, Formal analysis, Validation, Software, Writing - original draft. \textbf{Antonio Palermo:} Conceptualization, Software, Formal analysis, Writing - review \& editing, Supervision. \textbf{Alessandro Marzani:} Conceptualization, Writing - review \& editing, Supervision, Funding acquisition.

\section*{Declaration of competing interest}
\noindent The authors declare that they have no conflict of interest.

\section*{Acknowledgments}
\noindent This project has received funding from the European Union’s Horizon 2020 research and innovation programme under the Marie Skłodowska Curie grant agreement No 813424. 

\appendix

\section{Details on the transfer matrix method (TMM)}
\label{Appendix A}

In this Appendix, we provide the details of the transfer matrix method for the modulated beam (see Fig. \ref{fig:figA1}) \cite{chen2019nonreciprocal}. According to Euler beam theory, the $p$th order displacement in the $n$th cell can be expressed as:

\begin{equation} \label{equ:p-th disp}
    \hat{w}_n^{(p)}(x)=[\mathrm{e}^{\mathrm{i} \beta^{(p)}(x-x_{n})}, \mathrm{e}^{-\mathrm{i} \beta^{(p)}(x-x_{n})}, \mathrm{e}^{\beta^{(p)}(x-x_{n})}, \mathrm{e}^{-\beta^{(p)}(x-x_{n})}] [A_n^{(p)},B_n^{(p)},C_n^{(p)},D_n^{(p)}]^{T} := \mathcal{L}_n^{(p)}(x) \mathbf{U}_n^{(p)},
\end{equation}
\noindent
in which $\beta^{(p)}=\sqrt[4]{\rho \mathcal{A} (\omega+p \omega_m)^2/\mathcal{D}}$ is the $p$th order wavenumber.

\setcounter{figure}{0}
\begin{figure}[hbt!]
	\centering
	\includegraphics[width=0.80\textwidth]{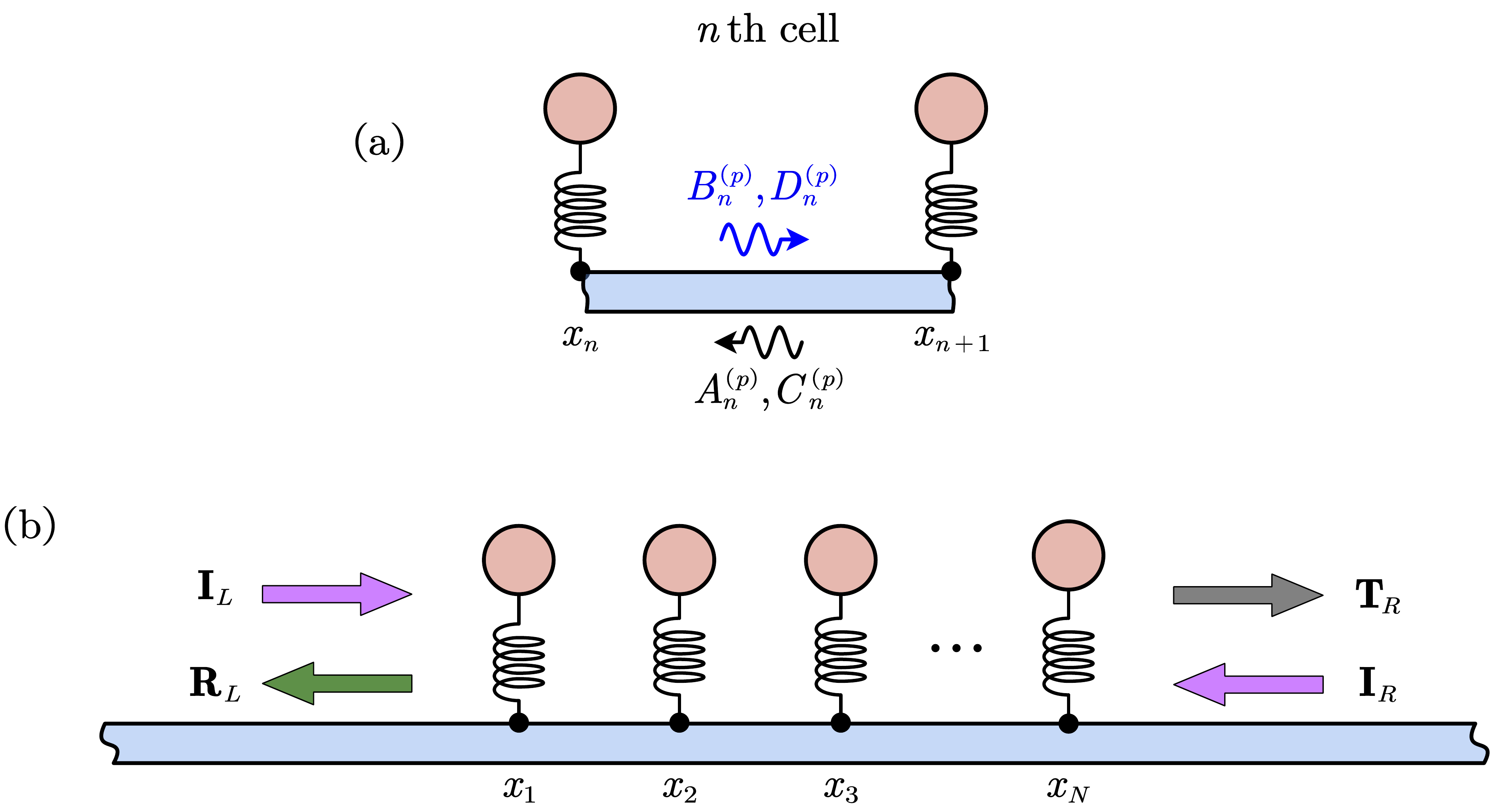}
	\caption{Schematic of transfer matrix method: (a) the $n$th cell, (b) the global system.}
	\label{fig:figA1}
\end{figure}

By truncating the orders from $p=-P$ to $p=P$, the displacement can be expressed in matrix form $\mathbf{\hat{w}}_n(x)=\mathbf{L}_n(x) \mathbf{A}_n$, with:

\begin{equation}
    \mathbf{\hat{w}}_n(x)=\left[\begin{array}{c}
    \hat{w}^{(-P)}(x) \\
    \hat{w}^{(-P+1)}(x) \\
    \vdots \\
    \hat{w}^{(P)}(x)
    \end{array}\right], \;
    \mathbf{L}_n(x)=\left[\begin{array}{cccc}
    \mathcal{L}_n^{(-P)}(x) & \mathbf{0} & \cdots & \mathbf{0} \\
    \mathbf{0} & \mathcal{L}_n^{(-P+1)}(x) & \cdots & \mathbf{0} \\
    \vdots & \vdots & \ddots & \vdots \\
    \mathbf{0} & \mathbf{0} & \cdots & \mathcal{L}_n^{(P)}(x)
    \end{array}\right], \;
    \mathbf{A}_n=\left[\begin{array}{c}
    \mathbf{U}_n^{(-P)} \\
    \mathbf{U}_n^{(-P+1)} \\
    \vdots \\
    \mathbf{U}_n^{(P)}
    \end{array}\right].
\end{equation}

Hence, the vertical force $\mathbf{\hat{F}}_n$ in Eq. \eqref{equ:force} can be written as:

\begin{equation} \label{equ:TMM force}
    \mathbf{\hat{F}}_n = \mathbf{D}_n \mathbf{M}_n^{-1} \mathbf{Q}_n \mathbf{\hat{w}}_n(x_n) = \mathbf{D}_n \mathbf{M}_n^{-1} \mathbf{Q}_n \mathbf{L}_n(x_n) \mathbf{A}_n.
\end{equation}

For an arbitrary $p$th order, the continuities of the displacement, slope, bending moment and shear force at $x_n$ yield:

\begin{subequations}
\begin{equation}  \label{equ:continuity of disp}
    \hat{w}_{n-1}^{(p)}(x_n) = \hat{w}_{n}^{(p)}(x_n), 
\end{equation}
\begin{equation} \label{equ:continuity of slope}
    \frac{\partial}{\partial x}\hat{w}_{n-1}^{(p)}(x_n) =  \frac{\partial}{\partial x}\hat{w}_{n}^{(p)}(x_n),
\end{equation}
\begin{equation} \label{equ:continuity of moment}
    \mathcal{D}\frac{\partial^2}{\partial x^2}\hat{w}_{n-1}^{(p)}(x_n) =  \mathcal{D}\frac{\partial^2}{\partial x^2}\hat{w}_{n}^{(p)}(x_n), 
\end{equation}
\begin{equation} \label{equ:continuity of force}
    \mathcal{D}\frac{\partial^3}{\partial x^3}\hat{w}_{n-1}^{(p)}(x_n) =  \mathcal{D}\frac{\partial^3}{\partial x^3}\hat{w}_{n}^{(p)}(x_n)-\hat{F}_{n}^{(p)}. 
\end{equation}
\end{subequations}

Substituting Eq. \eqref{equ:p-th disp} into Eqs. \eqref{equ:continuity of disp}-\eqref{equ:continuity of force} yields:

\begin{equation} \label{equ:disp relation}
    \boldsymbol\alpha_{n-1}^{(p)} \mathbf{U}_{n-1}^{(p)}=\boldsymbol\zeta_{n}^{(p)} \mathbf{U}_{n}^{(p)}+\boldsymbol\gamma_{n}^{(p)},
\end{equation}

\noindent with the coefficients:

\begin{equation}
    \boldsymbol\alpha_{n-1}^{(p)}=\left[\begin{array}{cccc}
    \mathrm{e}^{\mathrm{i}\beta^{(p)}\ell_n} & \mathrm{e}^{-\mathrm{i}\beta^{(p)}\ell_n} & \mathrm{e}^{\beta^{(p)}\ell_n} & \mathrm{e}^{-\beta^{(p)}\ell_n} \\
    \mathrm{i}\mathrm{e}^{\mathrm{i}\beta^{(p)}\ell_n} & -\mathrm{i}\mathrm{e}^{-\mathrm{i}\beta^{(p)}\ell_n} & \mathrm{e}^{\beta^{(p)}\ell_n} & -\mathrm{e}^{-\beta^{(p)}\ell_n} \\
    -\mathrm{e}^{\mathrm{i}\beta^{(p)}\ell_n} & -\mathrm{e}^{-\mathrm{i}\beta^{(p)}\ell_n} & \mathrm{e}^{\beta^{(p)}\ell_n} & \mathrm{e}^{-\beta^{(p)}\ell_n} \\
    -\mathrm{i}\mathrm{e}^{\mathrm{i}\beta^{(p)}\ell_n} & \mathrm{i}\mathrm{e}^{-\mathrm{i}\beta^{(p)}\ell_n} & \mathrm{e}^{\beta^{(p)}\ell_n} & -\mathrm{e}^{-\beta^{(p)}\ell_n}
    \end{array}\right], \;
    \boldsymbol\zeta_{n}^{(p)}=\left[\begin{array}{cccc}
    1 & 1 & 1 & 1 \\
    \mathrm{i} & -\mathrm{i} & 1 & -1 \\
    -1 & -1 & 1 & 1 \\
    -\mathrm{i} & \mathrm{i} & 1 & -1
    \end{array}\right], \;
    \boldsymbol\gamma_{n}^{(p)}=\left[\begin{array}{c}
    0 \\
    0 \\
    0 \\
    \hat{F}_{n}^{(p)}\chi^{(p)}
    \end{array}\right],
\end{equation}

\noindent where $\ell_n=x_n-x_{n-1}$, and $\chi^{(p)}=-1/[(\beta^{(p)})^3\mathcal{D}]$. Similarly, by truncating the orders from $-P$ to $P$, the displacements can be expressed in matrix form:

\begin{equation} \label{equ:TMM P-disp}
    \boldsymbol\alpha \mathbf{A}_{n-1}=\boldsymbol\zeta \mathbf{A}_{n}+\boldsymbol\gamma \hat{\mathbf{F}}_n,
\end{equation}

\noindent with:

\begin{equation}
\boldsymbol\alpha=\left[\begin{array}{cccc}
\boldsymbol\alpha_{n-1}^{(-P)} & & &\\
& \boldsymbol\alpha_{n-1}^{(-P+1)} & &\\
& & \ddots & \\
& & & \boldsymbol\alpha_{n-1}^{(P)}
\end{array}\right],\;
\boldsymbol\zeta=\left[\begin{array}{cccc}
\boldsymbol\zeta_{n}^{(-P)} & & &\\
& \boldsymbol\zeta_{n}^{(-P+1)} & &\\
& & \ddots & \\
& & & \boldsymbol\zeta_{n}^{(P)}
\end{array}\right],\; 
\boldsymbol\gamma=\left[\begin{array}{llll}
0 & 0 & \cdots & 0 \\
0 & 0 & \cdots & 0 \\
0 & 0 & \cdots & 0 \\
\chi^{(-P)} & 0 & \cdots & 0 \\
0 & 0 & \cdots & 0 \\
0 & 0 & \cdots & 0 \\
0 & 0 & \cdots & 0 \\
0 & \chi^{(-P+1)} & \cdots & 0 \\
\vdots & \vdots & \ddots & \vdots \\
0 & 0 & \cdots & 0 \\
0 & 0 & \cdots & 0 \\
0 & 0 & \cdots & 0 \\
0 & 0 & \cdots & \chi^{(P)}
\end{array}\right]
\end{equation}

Combining Eq. \eqref{equ:TMM force} and Eq. \eqref{equ:TMM P-disp} we obtain:

\begin{equation} \label{equ:TMM disp relation}
    \boldsymbol\alpha \mathbf{A}_{n-1} = \boldsymbol\zeta \mathbf{A}_{n}+\boldsymbol\gamma \mathbf{D}_n \mathbf{M}_n^{-1} \mathbf{Q}_n \mathbf{L}_n(x_n) \mathbf{A}_n,
\end{equation}

\noindent from which we obtain the local transfer matrix relating $\mathbf{A}_{n-1}$ to $\mathbf{A}_{n}$:

\begin{equation}
    \mathbf{T}_{n} \mathbf{A}_{n-1} = \mathbf{A}_{n},
\end{equation}

\noindent where:

\begin{equation}
    \mathbf{T}_{n} = [\boldsymbol\zeta+\boldsymbol\gamma \mathbf{D}_n \mathbf{M}_n^{-1} \mathbf{Q}_n \mathbf{L}_n(x_n)]^{-1}\boldsymbol\alpha.
\end{equation}

Therefore, for an infinite beam coupled with $N$ resonators, the global equation is expressed as:

\begin{equation} \label{equ:TMM global eq}
    \mathcal{T} \mathbf{A}_{0}= \mathbf{A}_{N},
\end{equation}

\noindent where the global transfer matrix $\mathcal{T}$ reads: 

\begin{equation}
    \mathcal{T} = \mathbf{T}_{N}\mathbf{T}_{N-1}\cdots\mathbf{T}_{1}.
\end{equation}

After some algebraic operations, Eq. \eqref{equ:TMM global eq} can be further written as:

\begin{equation} \label{equ:TMM global matrix}
    \left[\begin{array}{cc}
    \mathcal{M}_{11} & \mathcal{M}_{12} \\
    \mathcal{M}_{21} & \mathcal{M}_{22}
    \end{array}\right] \left[\begin{array}{c}
    \mathbf{I}_{L} \\
    \mathbf{R}_{L}
    \end{array}\right] = \left[\begin{array}{c}
    \mathbf{T}_{R} \\
    \mathbf{I}_{R}
    \end{array}\right],
\end{equation}

\noindent with coefficients:

\begin{subequations}
\begin{equation}
    \mathcal{M}=\mathcal{P}\mathcal{T}\mathcal{P}^T, 
\end{equation}
\begin{equation}
    \mathbf{I}_{L}=[B_0^{(-P)},D_0^{(-P)},B_0^{(-P+1)},D_0^{(-P+1)},\cdots,B_0^{(P)},D_0^{(P)}]^T, 
\end{equation}
\begin{equation}
    \mathbf{R}_{L}=[A_0^{(-P)},C_0^{(-P)},A_0^{(-P+1)},C_0^{(-P+1)},\cdots,A_0^{(P)},C_0^{(P)}]^T, 
\end{equation}
\begin{equation}
    \mathbf{T}_{R}=[B_N^{(-P)},D_N^{(-P)},B_N^{(-P+1)},D_N^{(-P+1)},\cdots,B_N^{(P)},D_N^{(P)}]^T, 
\end{equation}
\begin{equation}
    \mathbf{I}_{R}=[A_N^{(-P)},C_N^{(-P)},A_N^{(-P+1)},C_N^{(-P+1)},\cdots,A_N^{(P)},C_N^{(P)}]^T, 
\end{equation}
\end{subequations}

\noindent in which $\mathcal{P}$ is an elementary matrix which reads:

\begin{equation}
    \mathcal{P}=\left[\begin{array}{ccccccccc}
    0 & 1 & 0 & 0 & 0 & 0 & \cdots & 0 & 0 \\
    0 & 0 & 0 & 1 & 0 & 0 & \cdots & 0 & 0 \\
    0 & 0 & 0 & 0 & 0 & 1 & \cdots & 0 & 0 \\
    \vdots & \vdots & \vdots & \vdots & \vdots & \vdots & \ddots & \vdots & \vdots \\
    0 & 0 & 0 & 0 & 0 & 0 & \cdots & 0 & 1 \\
    1 & 0 & 0 & 0 & 0 & 0 & \cdots & 0 & 0 \\
    0 & 0 & 1 & 0 & 0 & 0 & \cdots & 0 & 0 \\
    0 & 0 & 0 & 0 & 1 & 0 & \cdots & 0 & 0 \\
    \vdots & \vdots & \vdots & \vdots & \vdots & \vdots & \ddots & \vdots & \vdots \\
    0 & 0 & 0 & 0 & 0 & 0 & \cdots & 1 & 0
    \end{array}\right]. 
\end{equation}

Eq. \eqref{equ:TMM global matrix} can be further transformed to:

\begin{equation} \label{equ:TMM global scattering}
    \boldsymbol{\psi}_\text{out}=\mathcal{S} \boldsymbol{\psi}_\text{in},
\end{equation}
where:
\begin{equation}
    \boldsymbol{\psi}_\text{out}=
    \left(\begin{array}{l}
    \mathbf{R}_{L} \\
    \mathbf{T}_{R}
\end{array}\right), \quad 
\boldsymbol{\psi}_\text{in}=
\left(\begin{array}{l}
    \mathbf{I}_{L} \\
    \mathbf{I}_{R}
\end{array}\right), \quad \mathcal{S}=\left(\begin{array}{cc}
    -\mathcal{M}_{22}^{-1}\mathcal{M}_{21} & \mathcal{M}_{22}^{-1} \\
    \mathcal{M}_{11}-\mathcal{M}_{12}\mathcal{M}_{22}^{-1}\mathcal{M}_{21} & \mathcal{M}_{12}\mathcal{M}_{22}^{-1}
\end{array}\right).
\end{equation}

With Eq. \eqref{equ:TMM global scattering} we can compute both the transmission and reflection coefficients directly. It is worth mentioning that, due to the presence of exponential amplification terms in $\boldsymbol\alpha_{n-1}^{(p)}$ in Eq. \eqref{equ:disp relation}, the transfer matrix method may encounter numerical divergence in some occasions, e.g., when considering a large number of oscillators or large values of resonators spacing. Such a limitation can be well addressed by the multiple scattering formulation proposed in this work.

\section{Validation of non-modulated dispersion equation} \label{Appendix B}

In this Appendix, we validate the analytical dispersion equation of non-modulated metamaterials via the finite element method (FEM). To do so, we build 2D FE models (unit cells) using 2D elasticity in COMSOL Multiphysics. In particular, the Euler beam is modeled by a 2D plane-stress FE model with dimensions $a\times h_b$ (Fig. \ref{fig:figB1}a), while the half-space is modeled by a 2D plane-strain FE model with the height $\ell_z=4\pi c_T/\omega_0$ (Fig. \ref{fig:figB1}b). To model the linear spring, we use a truss model with the unit cross-sectional area and unit height whose equivalent Young modulus satisfies $E_t=m_0\omega_0^2$. Additionally, the resonator mass is modeled by a point mass model with mass $m_0$. To simulate the dynamics of an infinite array of periodic resonators, we impose a pair of Floquet periodic boundary conditions on the vertical substrate edges. In Fig. \ref{fig:figB1}b, a clamped boundary condition is enforced at the bottom edge to avoid rigid motions. 

For the metabeam, the parameters used in this work are set as: mass density $\rho=2700$ kg/m$^3$, Young modulus $E=69$ GPa, Poisson ratio $\nu=0.33$, lattice constant $a=0.04$ m, beam thickness $h_b=0.002$ m, beam width $b_w=0.03$ m, resonance frequency of oscillators $\omega_0=80\pi$ rad/s, and damping coefficient of oscillators $c=0$. For the metasurface, the parameters used are: mass density $\rho=2700$ kg/m$^3$, Young modulus $E=69$ GPa, Poisson ratio $\nu=0.33$, lattice constant $a=0.3$ m, resonance frequency of oscillators $\omega_0=200\pi$ rad/s, and damping coefficient of oscillators $c=0$. The numerical dispersion curves are obtained by solving the eigenvalue problem for given wave number varying between $k=[0, \pi/a]$. The comparison between the analytical dispersion curves computed by Eqs. \eqref{equ:original dispersion equation of metabeam}, \eqref{equ:original dispersion equation of metasurface} and FE simulations for non-modulated metabeam and metasurface is shown in Fig. \ref{fig:figB1}c and Fig. \ref{fig:figB1}d, respectively. Excellent agreement between them is observed.

\setcounter{figure}{0}
\begin{figure}[hbt!]
	\centering
	\includegraphics[width=1\textwidth]{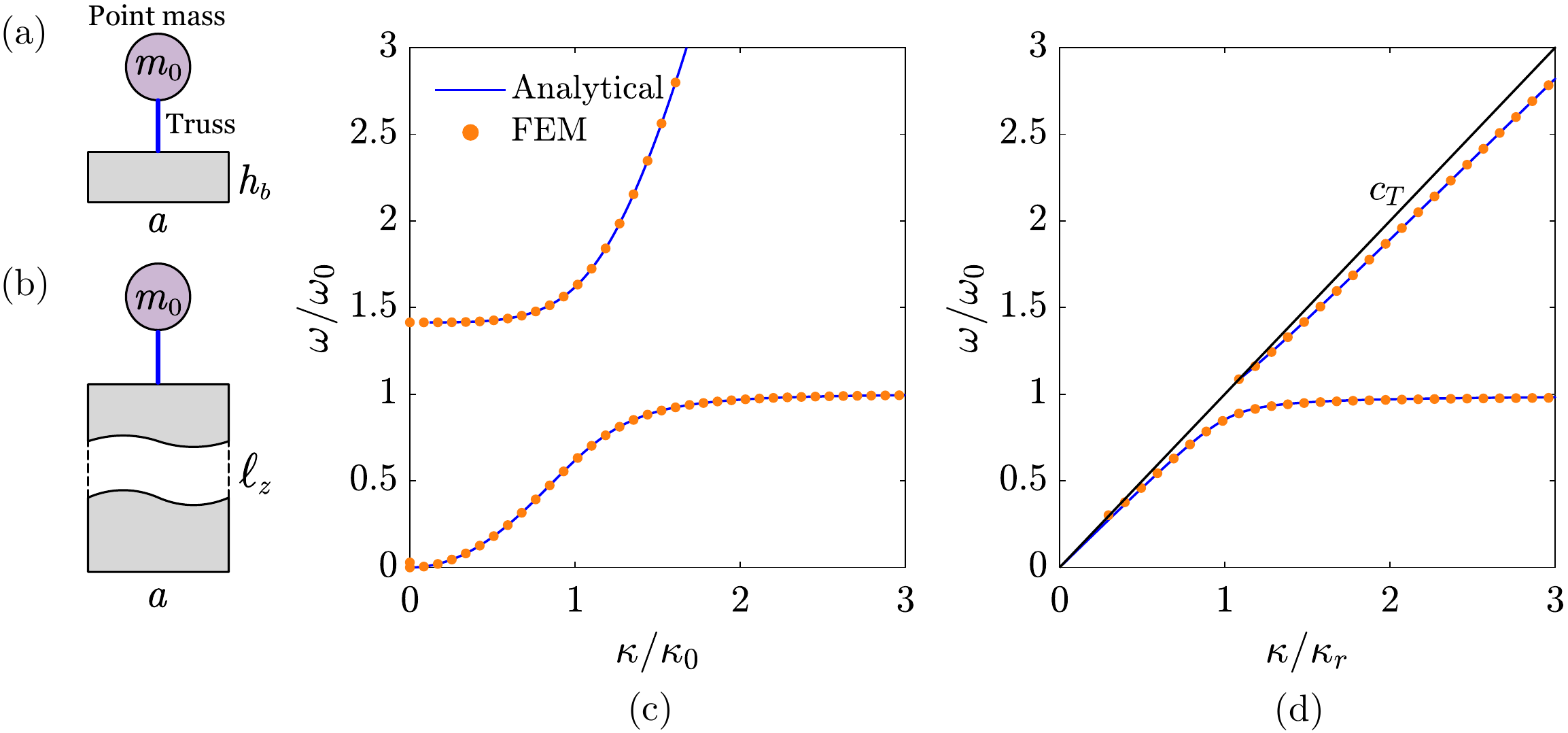}
	\caption{Comparison of non-modulated (original) dispersion curves between the analytical solution and the FE solution. (a, b) Schematic of unit cells used for the FE simulation. The dispersion curves for (c) flexural waves in a metabeam, and (d) Rayleigh waves in a metasurface.}
	\label{fig:figB1}
\end{figure}

\section{Details on the FE model for transient simulations} \label{Appendix C}

In this Appendix, we provide the details of the 2D plane-strain FE model, implemented in COMSOL Multiphysics, used to verify our analytical solutions in Section \ref{Steady-state solutions of SAW}. The FE model consists of an array of resonators and a substrate with width $\ell_x=32\lambda_0$ and depth $\ell_x=8\lambda_0$, where $\lambda_0=2\pi c_T/\omega_0$ (see Figs. \ref{fig:figC1}a,b). As in \ref{Appendix B}, the resonator is modeled by a point
mass $m_0$, while the spring is modeled by a truss element with unit length and cross-sectional area whose Young modulus reads $E_t =k_{0}+k_a\cos(\omega_m t-\kappa_m x)$. To minimize reflections from the domain borders, we add low-reflecting boundary conditions around the substrate (denoted by the dashed lines). The substrate is discretized using a fine mesh ($\lambda_0/10$) of quadratic serendipity elements, which allows to obtain convergent results at the frequency of interest.

We perform numerical simulations in the time domain. A narrow tone-burst signal of the form $F_0(t) = A_0 [H(t)-H(t-2\pi N/\omega)]\sin(\omega t)[1-\cos(\omega t/N)]$ is used to generate Rayleigh waves, where $H(t)$ is the Heaviside function. In the numerical example, the amplitude is set as $A_0=1$, the central frequency is $\omega=1.185\omega_0$, and the number of cycles is $N=60$. We display the signal and its Fourier spectrum in Figs. \ref{fig:figC1}c,d.

\setcounter{figure}{0}
\begin{figure}[hbt!]
	\centering
	\includegraphics[width=0.9\textwidth]{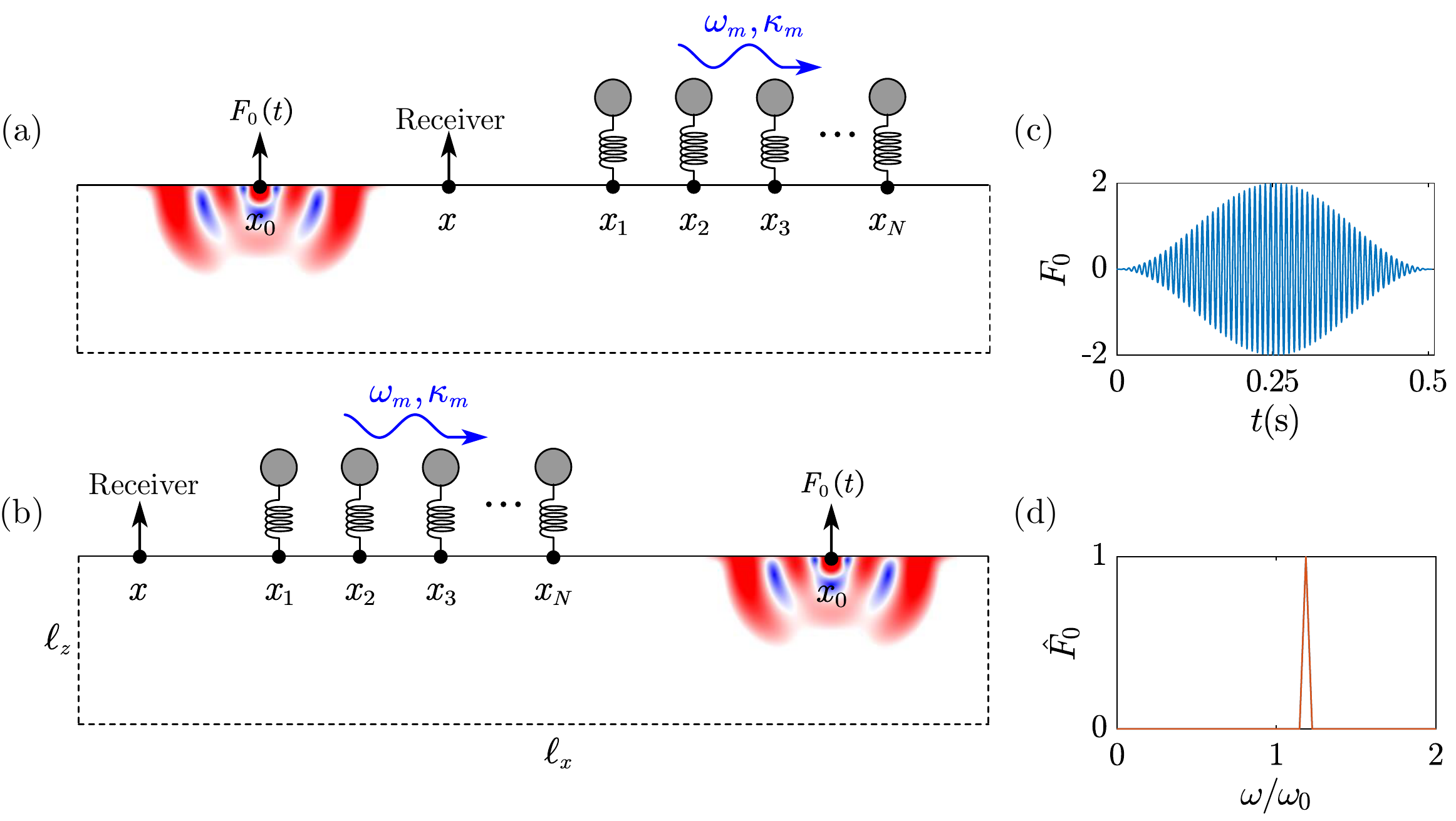}
	\caption{Schematic of the FE model for transient simulations.}
	\label{fig:figC1}
\end{figure}

\bibliography{mybibfile}

\end{document}